\documentstyle[12pt]{article}

\input{tcilatex}

\begin{document}

\title{Anholonomic Triads and New Classes of (2+1)--Dimensional Black Hole
Solutions}
\author{Sergiu I. Vacaru\thanks{%
e--mail:\ vacaru@fisica.ist.utl.pt, sergiu$_{-}$vacaru@yahoo.com}\ ,
Panyiotis Stavrinos \thanks{%
e--mail:\ pstavrin@cc.uoa.gr}\quad and Evghenii Gaburov \thanks{
e--mail:\ eg35@leicester.ac.uk } \quad \\
\\
{\small * \ Physics Department, CSU Fresno,\ Fresno, CA 93740-8031, USA, }\\
{\small \& }\\
{\small Centro Multidisciplinar de Astrofisica - CENTRA, Departamento de
Fisica,}\\
{\small Instituto Superior Tecnico, Av. Rovisco Pais 1, Lisboa, 1049-001,
Portugal,}\\
{\small {---}} \\
{\small $\dag$ Department of Mathematics, University of Athens,}\\
{\small 15784 Panepistimiopolis, Athens, Greece} \\
and \\
{\small $\ddag$ \ Department of Physics and Astronomy, University}\\
{\small of Leicester, University Road, Leicester, LE1 7RH, UK }}
\date{May 31, 2002}
\maketitle

\begin{abstract}
We apply the method of moving anholonomic frames in order to construct new
classes of solutions of the Einstein equations on (2+1)--dimensional
pseudo--Riemannian spaces. The anholonomy associated to a class of
off--diagonal metrics results in alternative classes of black hole solutions
which are constructed following distinguished (by nonlinear connection
structure) linear connections and metrics. There are investigated black
holes with deformed horizons and renormalized locally an\-iso\-trop\-ic
constants. We speculate on properties of such anisotropic black holes with
characteristics defined by anholonomic frames and anisotropic interactions
of matter and gravity. The thermodynamics of locally anisotropic black holes
is discussed in connection with a possible statistical mechanics background
based on locally anisotropic variants of Chern--Simons theories.
\end{abstract}



\vspace{0.5cm} 

\section{Introduction}

In recent years there has occurred a substantial interest to the
(2+1)--dimensional gravity and black holes and possible connections of such
objects with string/M--theory. Since the first works of Deser, Jackiv and 't
Hooft \cite{djh} and Witten \cite{w} on three dimensional gravity and the
seminal solution for (2+1)--black holes constructed by Ba\~nados,
Teitelboim, and Zanelli (BTZ) \cite{btz} the gravitational models in three
dimensions have become a very powerful tool for exploring the foundations of
classical and quantum gravity, black hole physics, as well the geometrical
properties of the spaces on which the low--dimensional physics takes place %
\cite{cm}.

On the other hand, the low--dimensional geometries could be considered as an
arena for elaboration of new methods of solution of gravitational field
equations. One of peculiar features of general relativity in 2+1 dimensions
is that the bulk of physical solutions of Einstein equations are constructed
for a negative cosmological constant and on a space of constant curvature.
There are not such limitations if anholonomic frames modeling locally
anisotropic (la) interactions of gravity and matter are considered.

In our recent works \cite{v3} we emphasized the importance of definition of
frames of reference in general relativity in connection with new methods of
construction of solutions of the Einstein equations. The former priority
given to holonomic frames holds good for the 'simplest' spherical symmetries
and is less suitable for construction of solutions with 'deformed'
symmetries, for instance, of static black holes with elliptic (or
ellipsoidal and/or torus) configurations of horizons. Such type of
'deformed', locally anisotropc, solutions of the Einstein equations are
easily to be derived from the ansatz of metrics diagonalized with respect to
some classes of anholonomic frames induced by locally anisotropic
'elongations' of partial derivatives. After the task has been solved in
anholonomic variables it can be removed with respect to usual coordinate
bases when the metric becomes off--diagonal and the (for instance, elliptic)
symmetry is hiden in some nonlinear dependencies of the metric components.

The specific goal of the present work is to formulate the (2+1)--dimensional
gravity theory with respect to anholonomic frames with associated nonlinear
connection (N--connection) structure and to construct and investigate some
new classes of solutions of Einstein equations on locally anisotropic
spacetimes (modelled as usual pseudo--Riemannian spaces provided with an
anholonomic frame structure). A material of interest are the properties of
the locally anisotropic elastic media and rotating null fluid and
anisotropic collapse described by gravitational field equations with locally
anisotropic matter. We investigate black hole solutions that arise from
coupling in a self--consistent manner the three dimensional (3D)
pseudo--Riemannian geometry to the physics of locally anisotropic fluids
formulated with respect to anholonomic frames of reference. For certain
special cases the locally anisotropic matter gives the BTZ black holes
with/or not rotation and electrical charge and variants of their anisotropic
generalizations. For other cases, the resulting solutions are generic black
holes with ''locally anisotropic hair''.

We emphasize that the anisotropic gravitational field has very unusual
properties. For instance, the vacuum solutions of Einstein anisotropic
gravitational field equations could describe anisotropic black holes with
elliptic symmetry. Some subclasses of such locally anisotropic spaces are
teleparallel (with non--zero induced torsion but with vanishing curvature
tensor) another are characterized by nontrivial, induced from general
relativity on anholonomic frame bundle, N--connection and Riemannian
curvature and anholonomy induced torsion. In a more general approach the
N--connection and torsion are induced also from the condition that metric
and nonlinear connection must solve the Einstein equations.

The paper is organized as follows: In the next section we briefly review the
locally anisotropic gravity in (2+1)--dimensions. Conformal transforms with
anisotropic factors and corresponding classes of solutions of Einstein
equations with dynamical equations for N--connection coefficients are
examined in Sec. 3. In Sec. 4 we derive the energy--momentum tensors for
locally an\-iso\-tro\-pic elastic media and rotating null fluids. Sec. 5 is
devoted to the local anisotropy of (2+1)--dimensional solutions of Einstein
equations with anisotropic matter. The nonlinear self--polarization of
anisotropic vacuum gravitational fields and matter induced polarizations and
related topics on anisotropic black hole solutions are considered in Sec. 6.
We derive some basic formulas for thermodynamics of anisotropic black holes
in Sec. 7. The next Sec. 8 provides a statistical mechanics background for
locally anisotropic thermodynamics starting from the locally anisotropic
variants of Chern--Simons and Wess--Zumino-Witten models of locally
anisotropic gravity. Finally, in Sec. 9 we conclude and discuss the obtained
results.

\section{Anholonomic Frames and 3D Gravity}

In this Section we wish to briefly review and reformulate the Cartan's
method of moving frames \cite{cartan1} for investigation of gravitational
and matter field interactions with mixed subsets of holonomic
(unconstrained) and anholonomic (constrained, equivalently, locally
anisotropic, in brief, la) variables \cite{v3}. Usual tetradic (frame, or
vielbein) approaches to general relativity, see, for instance, \cite{mtw,haw}%
, consider 'non--mixed' cases when all basic vectors are anholonomic or
transformed into coordinate (holonomic) ones. We note that a more general
geometric background for locally anisotropic interactions and locally
anisotropic spacetimes, with applications in physics, was elaborated by
Miron and Anastasiei \cite{ma} in their generalized Finsler and Lagrange
geometry; further developments for locally anisotropic spinor bundles and
locally anisotropic superspaces are contained in Refs \cite{v1,v2}. Here we
restrict our constructions only to three dimensional (3D) pseudo--Riemannian
spacetimes provided with a global splitting characterized by two holonomic
and one anholonomic coordinates.

\subsection{Anholonomic frames and nonlinear connections}

We model the low dimensional spacetimes as a smooth (i. e. class $C^\infty )$
3D (pseudo) Riemannian manifolds $V^{(3)}$ being Hausdorff, paracompact and
connected and enabled with the fundamental structures of symmetric metric $%
g_{\alpha \beta },$ with signature $(-,+,+)$ and of linear, in general
nonsymmetric (if we consider anholonomic frames), metric connection $\Gamma
_{~\beta \gamma }^\alpha $ defining the covariant derivation $D_\alpha .$
The indices of geometrical objects on $V^{(3)}$ are stated with respect to a
frame vector field (triad, or dreibien) $e_\alpha $ and its dual $e^\alpha .$
A holonomic frame structure on 3D spacetime could be given by a local
coordinate base 
\begin{equation}  \label{pder}
\partial _\alpha =\partial /\partial u^\alpha ,
\end{equation}
consisting from usual partial derivatives on local coordinates $u=\{u^\alpha
\}$ and the dual basis 
\begin{equation}  \label{pdif}
d^\alpha =du^\alpha ,
\end{equation}
consisting from usual coordinate differentials $du^\alpha .$

An arbitrary holonomic frame $e_\alpha $ could be related to a coordinate
one by a local linear transform $e_\alpha =A_\alpha ^{~\beta }(u) \partial
_\beta ,$ for which the matrix $A_\alpha ^{~\beta }$ is nondegenerate and
there are satisfied the holonomy conditions 
\[
e_\alpha e_\beta -e_\beta e_\alpha =0. 
\]

Let us consider a 3D metric parametrized into $\left( 2+1\right) $
components 
\begin{equation}  \label{ansatz}
g_{\alpha \beta }=\left[ 
\begin{array}{cc}
g_{ij}+N_i^{\bullet }N_j^{\bullet }h_{\bullet \bullet } & N_j^{\bullet
}h_{\bullet \bullet } \\ 
N_i^{\bullet }h_{\bullet \bullet } & h_{\bullet \bullet }%
\end{array}
\right]
\end{equation}
given with respect to a local coordinate basis (\ref{pdif}), $du^\alpha
=\left( dx^i,dy\right) ,$ where the Greek indices run values $1,2,3,$ the
Latin indices $i,j,k,...$ from the middle of the alphabet run values for $%
n=1,2,...$ and the Latin indices from the beginning of the alphabet, $%
a,b,c,...,$ run values for $m=3,4,....$ if we wont to consider imbeddings of
3D spaces into higher dimension ones. The coordinates $x^i$ are treated as
isotropic ones and the coordinate $y^{\bullet }=y$ is considered anholonomic
(anisotropic). For 3D we denote that $a,b,c,...=\bullet ,$ $y^{\bullet
}\rightarrow y,$ $h_{ab}\rightarrow h_{\bullet \bullet }=h$ and $%
N_i^a\rightarrow N_i^{\bullet }=w_i.$ The coefficients $g_{ij}=g_{ij}\left(
u\right) ,h_{\bullet \bullet }=h\left( u\right) $ and $N_i^{\bullet }=N_i(u)$
are supposed to solve the 3D Einstein gravitational field equations. The
metric (\ref{ansatz}) can be rewritten in a block $(2\times 2)\oplus 1$ form 
\begin{equation}  \label{dm}
g_{\alpha \beta }=\left( 
\begin{array}{cc}
g_{ij}(u) & 0 \\ 
0 & h(u)%
\end{array}
\right)
\end{equation}
with respect to the anholonomic basis (frame, anisotropic basis) 
\begin{equation}  \label{dder}
\delta _\alpha = (\delta _i,\partial _{\bullet }) = \frac \delta {\partial
u^\alpha } = \left( \delta _i=\frac \delta {\partial x^i}=\frac \partial
{\partial x^i}-N_i^{\bullet }\left( u\right) \frac \partial {\partial
y},\partial _{\bullet }=\frac \partial {\partial y}\right)
\end{equation}
and its dual anholonomic frame 
\begin{equation}  \label{ddif}
\delta ^\beta = \left( d^i,\delta ^{\bullet }\right) =\delta u^\beta =
\left( d^i=dx^i,\delta ^{\bullet }=\delta y=dy+N_k^{\bullet }\left( u\right)
dx^k\right) .  \nonumber
\end{equation}
where the coefficients $N_j^{\bullet }\left( u\right) $ from (\ref{dder})
and (\ref{ddif}) could be treated as the components of an associated
nonlinear connection (N--connection) structure \cite{barth,ma,v1,v2} which
was considered in Finsler and generalized Lagrange geometries and applied in
general relativity and Kaluza--Klein gravity for construction of new classes
of solutions of Einstein equations by using the method of moving anholonomic
frames \cite{v3}. On 3D (pseudo)--Riemannian spaces the coefficients $%
N_j^{\bullet}$ define a triad of basis vectors (dreibein) with respect to
which the geometrical objects (tensors, connections and spinors) are
decomposed into holonomic (with indices $i,j,...$) and anholonomic (provided
with $\bullet$--index) components.

A local frame (local basis) structure $\delta _\alpha $ on $V^{(3)}\to
V^{(2+1)}$ (by $(2+1)$ we denote the N--connection splitting into $2$
holonomic and $1$ anholonomic variables in explicit form; this decomposition
differs from the usual two space and one time--like parametrizations) is
characterized by its anholonomy coefficients $w_{~\beta \gamma }^\alpha $
defined from relations 
\begin{equation}  \label{anholon}
\delta _\alpha \delta _\beta -\delta _\beta \delta _\alpha =w_{~\alpha \beta
}^\gamma \delta _\gamma .
\end{equation}
The rigorous mathematical definition of N--connection is based on the
formalism of horizontal and vertical subbundles and on exact sequences in
vector bundles \cite{barth,ma}. In this work we introduce a N--connection as
a distribution which for every point $u=(x,y)\in V^{(2+1)}$ defines a local
decomposition of the tangent space 
\[
T_uV^{(2+1)}=H_uV^{(2)}\oplus V_uV^{(1)}. 
\]
into horizontal subspace, $H_uV^{(2)},$ and vertical (an\-iso\-tro\-py)
subspace, $V_uV^{(1)},$ which is given by a set of coefficients $%
N_j^{\bullet }\left( u^\alpha \right) .$ A N--connection is characterized by
its curvature 
\begin{equation}  \label{ncurv}
\Omega _{ij}^{\bullet }=\frac{\partial N_i^{\bullet }}{\partial x^j}-\frac{%
\partial N_j^{\bullet }}{\partial x^i}+N_i^{\bullet }\frac{\partial
N_j^{\bullet }}{\partial y}-N_j^{\bullet }\frac{\partial N_i^{\bullet }}{%
\partial y}.
\end{equation}
The class of usual linear connections can be considered as a particular case
of N--connecti\-ons when 
\[
N_j^{\bullet }(x,y)=\Gamma _{\bullet j}^{\bullet }(x)y^{\bullet }. 
\]
The elongation (by N--connection) of partial derivatives and differentials
in the adapted to the N--connection operators (\ref{dder}) and (\ref{ddif})
reflects the fact that on the (pseudo) Riemannian spacetime $V^{(2+1)}$ it
is modeled a generic local anisotropy characterized by anholonomy relations (%
\ref{anholon}) when the anholonomy coefficients are computed as follows 
\begin{eqnarray}
w_{~ij}^k & = & 0,w_{~\bullet j}^k=0,w_{~i\bullet}^k=0, w_{~\bullet
\bullet}^k=0,w_{~\bullet \bullet}^\bullet=0,  \label{anhol} \\
w_{~ij}^\bullet & = & -\Omega _{ij}^\bullet, w_{~\bullet j}^\bullet =
-\partial _\bullet N_i^\bullet, w_{~i\bullet }^\bullet = \partial _\bullet
N_i^\bullet.  \nonumber
\end{eqnarray}
The frames (\ref{dder}) and (\ref{ddif}) are locally adapted to the
N--connection structure, define a local anisotropy and, in brief, are called
anholonomic bases. A N--connection structure distinguishes (d) the
geometrical objects into horizontal and vertical components, i. e. transform
them into d--objects which are briefly called d--tensors, d--metrics and
d--connections. Their components are defined with respect to an anholonomic
basis of type (\ref{dder}), its dual (\ref{ddif}), or their tensor products
(d--linear or d--affine transforms of such frames could also be considered).
For instance, a covariant and contravariant d--tensor $Q,$ is expressed 
\begin{equation}
Q = Q_{~\beta }^\alpha \delta _\alpha \otimes \delta ^\beta = Q_{~j}^i\delta
_i\otimes d^j+ Q_{~\bullet }^i\delta _i\otimes \delta ^{\bullet} +
Q_{~j}^{\bullet }\partial _{\bullet }\otimes d^j+ Q_{~\bullet }^{\bullet}
\partial _{\bullet }\otimes \delta ^{\bullet }.  \nonumber
\end{equation}
Similar decompositions on holonomic--anholonomic, conventionally on
horizontal (h) and vertical (v) components, hold for connection, torsion and
curvature components adapted to the N--connection structure.

\subsection{Compatible N- and d--connections and metrics}

A linear d--connection $D$ on a locally anisotropic spacetime $V^{(2+1)},$ \ 
$D_{\delta _\gamma }\delta _\beta =\Gamma _{~\beta \gamma }^\alpha \left(
x,y\right) \delta _\alpha , $ is given by its h--v--components, 
\begin{equation}
\Gamma _{~\beta \gamma }^\alpha =\left( L_{~jk}^i,L_{~\bullet k}^{\bullet
},C_{~j\bullet }^i,C_{~\bullet \bullet }^{\bullet }\right)  \nonumber
\end{equation}
where 
\begin{equation}
D_{\delta _k}\delta _j = L_{~jk}^i\delta _i, D_{\delta _k}\partial _\bullet
= L_{\bullet k}^\bullet \partial _\bullet, D_{\partial _\bullet }\delta _j =
C_{~j\bullet }^i\delta _i, D_{\delta_\bullet }\partial _\bullet = C_{\bullet
\bullet }^\bullet \partial _\bullet.
\end{equation}
A metric on $V^{(2+1)}$ with its coefficients parametrized as (\ref{ansatz})
can be written in distinguished form (\ref{dm}), as a metric d--tensor (in
brief, d--metric), with respect to an anholonomic base (\ref{ddif}), i. e. 
\begin{equation}  \label{dmetric}
\delta s^2 = g_{\alpha \beta }\left( u\right) \delta ^\alpha \otimes \delta
^\beta = g_{ij}(x,y)dx^idx^j+h(x,y)(\delta y)^2.
\end{equation}
Some N--connection, d--connection and d--metric structures are compatible if
there are satisfied the conditions 
\[
D_\alpha g_{\beta \gamma }=0. 
\]
For instance, a canonical compatible d--connection 
\[
^c\Gamma _{~\beta \gamma }^\alpha =\left( ^cL_{~jk}^i,^cL_{~\bullet
k}^{\bullet },^cC_{~j\bullet }^i,^cC_{~\bullet \bullet }^{\bullet }\right) 
\]
is defined by the coefficients of d--metric (\ref{dmetric}), $g_{ij}\left(
x,y\right) $ and $h\left( x,y\right) ,$ and by the coefficients of
N--connection, 
\begin{eqnarray}
^cL_{~jk}^i & = & \frac 12g^{in}\left( \delta _kg_{nj}+ \delta
_jg_{nk}-\delta _ng_{jk}\right) ,  \nonumber \\
^cL_{~\bullet k}^\bullet & = & \partial _\bullet N_k^\bullet+\frac 12h^{-1}
\left( \delta _kh- 2 h \partial _\bullet N_i^\bullet \right) ,  \nonumber \\
^cC_{~j\bullet }^i & = & \frac 12g^{ik}\partial _\bullet g_{jk},  \nonumber
\\
^cC_{~\bullet \bullet}^\bullet & = & \frac 12h^{-1} \left( \partial _\bullet
h \right).  \label{dcon}
\end{eqnarray}
The coefficients of the canonical d--connection generalize for locally
anisotropic spacetimes the well known Christoffel symbols. By a local linear
non--degenerate transform to a coordinate frame we obtain the coefficients
of the usual (pseudo) Riemannian metric connection. For a canonical
d--connection (\ref{dcon}), hereafter we shall omit the left--up index
''c'', the components of canonical torsion, 
\begin{eqnarray}
&T\left( \delta _\gamma ,\delta _\beta \right) &= T_{~\beta \gamma }^\alpha
\delta _\alpha ,  \nonumber \\
&T_{~\beta \gamma }^\alpha &= \Gamma _{~\beta \gamma }^\alpha - \Gamma
_{~\gamma \beta }^\alpha +w_{~\beta \gamma }^\alpha  \nonumber
\end{eqnarray}
are expressed via d--torsions 
\begin{eqnarray}
T_{.jk}^i & = & T_{jk}^i=L_{jk}^i-L_{kj}^i,\quad
T_{j\bullet}^i=C_{.j\bullet}^i,T_{\bullet j}^i=-C_{j \bullet}^i,  \nonumber
\\
T_{.bc}^a &= &S_{.bc}^a=C_{bc}^a-C_{cb}^a\to S_{.\bullet \bullet}^\bullet
\equiv 0,  \label{dtorsions} \\
T_{.ij}^\bullet & = & -\Omega _{ij}^\bullet, \quad T_{.\bullet i}^\bullet =
\partial _\bullet N_i^\bullet -L_{.\bullet i}^\bullet ,\quad T_{.i\bullet
}^\bullet = -T_{.\bullet i}^\bullet  \nonumber
\end{eqnarray}
which reflects the anholonomy of the corresponding locally anisotropic frame
of reference on $V^{(2+1)};$ they are induced effectively. With respect to
holonomic frames the d--torsions vanishes. Putting the non--vanishing
coefficients (\ref{dcon}) into the formula for curvature 
\begin{eqnarray}
R\left( \delta _\tau ,\delta _\gamma \right) \delta _\beta &= & R_{\beta
~\gamma\tau }^{~\alpha }\delta _\alpha ,  \nonumber \\
R_{\beta ~\gamma \tau }^{~\alpha } & = & \delta _\tau \Gamma _{~\beta \gamma
}^\alpha -\delta _\gamma \Gamma _{~\beta \delta }^\alpha + \Gamma _{~\beta
\gamma }^\varphi \Gamma _{~\varphi \tau }^\alpha -\Gamma _{~\beta \tau
}^\varphi \Gamma _{~\varphi \gamma }^\alpha + \Gamma _{~\beta \varphi
}^\alpha w_{~\gamma \tau }^\varphi  \nonumber
\end{eqnarray}
we compute the components of canonical d--curvatures 
\begin{eqnarray}
R_{h.jk}^{.i} & = & \delta _kL_{.hj}^i-\delta_jL_{.hk}^i +
L_{.hj}^mL_{mk}^i- L_{.hk}^mL_{mj}^i-C_{.h\bullet }^i\Omega _{.jk}^\bullet, 
\nonumber \\
R_{\bullet .jk}^{.\bullet} & = & \delta _kL_{.\bullet j}^\bullet
-\delta_jL_{.\bullet k}^\bullet -C_{.\bullet \bullet}^\bullet \Omega
_{.jk}^\bullet,  \label{dcurvatures} \\
P_{j.k\bullet}^{.i} & = & \delta _k L_{.jk}^i + C_{.j\bullet
}^iT_{.k\bullet}^\bullet - ( \delta
_kC_{.j\bullet}^i+L_{.lk}^iC_{.j\bullet}^l - L_{.jk}^lC_{.l\bullet}^i -
L_{.\bullet k}^\bullet C_{.j\bullet}^i ),  \nonumber \\
P_{\bullet.k\bullet}^{.\bullet} & = & \partial _\bullet L_{.\bullet
k}^\bullet + C_{.\bullet \bullet}^\bullet T_{.k\bullet}^\bullet - ( \delta
_kC_{.\bullet \bullet }^\bullet - L_{.\bullet k}^\bullet C_{.\bullet \bullet
}^\bullet ),  \nonumber \\
S_{j.bc}^{.i} & = & \partial _cC_{.jb}^i-\partial _bC_{.jc}^i +
C_{.jb}^hC_{.hc}^i-C_{.jc}^hC_{hb}^i \to S_{j.\bullet \bullet}^{.i} \equiv 0
,  \nonumber \\
S_{b.cd}^{.a} & = &\partial _dC_{.bc}^a-\partial _cC_{.bd}^a +
C_{.bc}^eC_{.ed}^a-C_{.bd}^eC_{.ec}^a \to S_{\bullet.\bullet
\bullet}^{.\bullet} \equiv 0.  \nonumber
\end{eqnarray}
The h--v--decompositions for the torsion, (\ref{dtorsions}), and curvature, (%
\ref{dcurvatures}), are invariant under local coordinate transforms adapted
to a prescribed N--connection structure.

\subsection{Anholonomic constraints and Einstein equations}

The Ricci d--tensor $R_{\beta \gamma }=R_{\beta ~\gamma \alpha }^{~\alpha }$
has the components 
\begin{eqnarray}
R_{ij} & = & R_{i.jk}^{.k},\quad R_{i\bullet
}=-^2P_{ia}=-P_{i.k\bullet}^{.k},  \label{dricci} \\
R_{\bullet i} &= & ^1P_{\bullet i}= P_{\bullet .i\bullet}^{.\bullet},\quad
R_{ab}=S_{a.bc}^{.c}\to S_{\bullet \bullet} \equiv 0  \nonumber
\end{eqnarray}
and, in general, this d--tensor is non symmetric. We can compute the scalar
curvature $\overleftarrow{R}=g^{\beta \gamma }R_{\beta \gamma }$ of a
d-connection $D,$%
\begin{equation}  \label{dscalar}
{\overleftarrow{R}}=\widehat{R}+S,
\end{equation}
where $\widehat{R}=g^{ij}R_{ij}$ and $S=h^{ab}S_{ab}\equiv 0$ for one
dimensional anisotropies. By introducing the values (\ref{dricci}) and (\ref%
{dscalar}) into the usual Einstein equations 
\begin{equation}  \label{einst1}
G_{\alpha \beta }+\Lambda g_{\alpha \beta }=k\Upsilon _{\beta \gamma },
\end{equation}
where 
\begin{equation}  \label{einstdt}
G_{\alpha \beta }=R_{\beta \gamma }-\frac 12g_{\beta \gamma }R
\end{equation}
is the Einstein tensor, written with respect to an anholonomic frame of
reference, we obtain the system of field equations for locally anisotropic
gravity with N--connection structure \cite{ma}: 
\begin{eqnarray}
R_{ij}-\frac 12\left( \widehat{R} - 2\Lambda \right) g_{ij} & = & k\Upsilon
_{ij},  \label{einsteq2a} \\
-\frac 12\left( \widehat{R}-2\Lambda\right) h_{\bullet \bullet} & = &
k\Upsilon _{\bullet \bullet},  \label{einsteq2b} \\
^1P_{\bullet i} & = & k\Upsilon _{\bullet i},  \label{einsteq2c} \\
^2P_{i\bullet} & = & -k\Upsilon _{i\bullet},  \label{einsteq2d}
\end{eqnarray}
where $\Upsilon _{ij},\Upsilon _{\bullet \bullet },\Upsilon _{\bullet i}$
and $\Upsilon _{i\bullet }$ are the components of the energy--momentum
d--tensor field $\Upsilon _{\beta \gamma }$ which includes the cosmological
constant terms and possible contributions of d--torsions and matter, and $k$
is the coupling constant.

The bulk of nontrivial locally isotropic solutions in 3D gravity were
constructed by considering a cosmological constant $\Lambda =-1/l^2,$ with
and equivalent vacuum energy--momentum $\Upsilon _{\beta \gamma }^{(\Lambda
)}=-\Lambda g_{\beta \gamma }.$

\subsection{Some ansatz for d--metrics}

\subsubsection{Diagonal d-metrics}

Let us introduce on 3D locally anisotropic spacetime $V^{(2+1)}$ the local
coordinates\newline
$(x^1,x^2,y),$ where $y$ is considered as the anisotropy coordinate, and
parametrize the d--metric (\ref{dmetric}) in the form 
\begin{equation}  \label{dm2}
\delta s^2=a\left( x^i\right) \left( dx^1\right) ^2+b\left( x^i\right)
(dx^2)^2+h\left( x^i,y\right) (\delta y)^2,
\end{equation}
where 
\[
\delta y=dy + w_1(x^i,y)dx^1+w_2(x^i,y)dx^2, 
\]
i. e. $N_i^\bullet =w_i(x^i,y).$

With respect to the coordinate base (\ref{pder}) the d--metric (\ref{dmetric}%
) transforms into the ansatz 
\begin{equation}  \label{m2}
g_{\alpha \beta }=\left[ 
\begin{array}{ccc}
a+w_1^{\ 2}h & w_1w_2h & w_1h \\ 
w_1w_2h & b+w_2^{\ 2}h & w_2h \\ 
w_1h & w_2h & h%
\end{array}
\right] .
\end{equation}

The nontrivial components of the Ricci d--tensor (\ref{dricci}) are computed 
\begin{eqnarray}  \label{ricci1d}
2abR_1^1 &=& 2abR_2^2 -\ddot b+\frac 1{2b}\dot b^2+\frac 1{2a}\dot a\dot
b+\frac 1{2b}a^{\prime }b^{\prime }-a^{\prime \prime }+\frac 1{2a}(a^{\prime
})^2  \nonumber
\end{eqnarray}
where the partial derivatives are denoted, for instance, $\dot h=\partial
h/\partial x^1,h^{\prime }=\partial h/\partial x^2$ and $h^{*}=\partial
h/\partial y.$ The scalar curvature is $R=2R_1^1.$

The Einstein d--tensor has a nontrivial component 
\[
G_3^3=-hR_1^1. 
\]

In the vacuum case with $\Lambda =0,$ the Einstein equ\-a\-tions (\ref%
{einsteq2a})--(\ref{einsteq2d}) are satisfied by arbitrary functions $%
a\left(x^i\right) ,b\left( x^i\right) $ solving the equation 
\begin{equation}  \label{vaceq}
-\ddot b+\frac 1{2b}\dot b^2+\frac 1{2a}\dot a\dot b+\frac 1{2b}a^{\prime
}b^{\prime }-a^{\prime \prime }+\frac 1{2a}(a^{\prime })^2=0
\end{equation}
and arbitrary function $h\left( x^i,y\right).$ Such functions should be
defined following some boundary conditions in a manner as to have
compatibility with the locally isotropic limit.

\subsubsection{Off--diagonal d--metrics}

For our further investigations it is convenient to consider d--metrics of
type 
\begin{equation}  \label{dm2a}
\delta s^2=g\left( x^i\right) \left( dx^1\right) ^2+2dx^1dx^2+h\left(
x^i,y\right) (\delta y)^2.
\end{equation}
The nontrivial components of the Ricci d--tensor are 
\begin{equation}  \label{ricci2d}
R_{11} =\frac 12g\frac{\partial ^2g}{\partial (x^2)^2},\qquad R_{12}
=R_{21}=\frac 12\frac{\partial ^2g}{\partial (x^2)^2},
\end{equation}
when the scalar curvature is $R=2R_{12}$ and the nontrivial component of the
Einstein d--tensor is 
\[
G_{33}=-\frac h2\frac{\partial ^2g}{\partial (x^2)^2}. 
\]

We note that for the both d--metric ansatz (\ref{dm2}) and (\ref{dm2a}) and
corresponding coefficients of Ricci d--tensor, (\ref{ricci1d}) and (\ref%
{ricci2d}), the h--components of the Einstein d--tensor vanishes for
arbitrary values of metric coefficients, i. e. $G_{ij}=0.$ In absence of
matter such ansatz admit arbitrary nontrivial anholonomy (N--connection and
N--curvature) coefficients (\ref{anhol}) becouse the values $w_{i}$ are not
contained in the 3D vacuum Einstein equations. The h--component of the
d--metric, $h(x^{k},y),$ and the coefficients of d--connection, $%
w_{i}(x^{k},y),$ are to be defined by some boundary conditions (for
instance, by a compatibility with the locally isotropic limit) and
compatibility conditions between nontrivial values of the cosmological
constant and energy--momentum d--tensor.

\section{Conformal Transforms with Anisotropic Factors}

One of pecular proprieties of the d--metric ansatz (\ref{dm2}) and (\ref%
{dm2a}) is that there is only one non--trivial component of the Einstein
d--tensor, $G_{33}.$ Becouse the values $P_{3i}$ and $P_{i3}$ for the
equations (\ref{einsteq2b}) and (\ref{einsteq2c}) vanish identically the
coefficients of N--connection, $w_i,$ are not contained in the Einstein
equations and could take arbitrary values. For static anisotropic
configurations the solutions constructed in Sections IV and V cand be
considered as 3D black hole like objects embedded in a locally anisotropic
background with prescribed anholonomic frame (N--connection) structure.

In this Section we shall proof that there are d--metrics for which the
Einstein equations reduce to some dynamical equations for the N--connection
coefficients.

\subsection{Conformal transforms of d--metrics}

A conformal transform of a d--metric 
\begin{equation}  \label{conft}
\left( g_{ij},h_{ab}\right) \longrightarrow \left( \widetilde{g}_{ij}=\Omega
^2\left( x^i,y\right) g_{ij},\widetilde{h}_{ab}= \Omega ^2\left(x^i,y\right)
h_{ab}\right)
\end{equation}
with fixed N--connection structure, $\widetilde{N}_i^a=N_i^a,$ deforms the
coefficients of canonical d--connection, 
\[
\widetilde{\Gamma }_{\ \beta \gamma }^\alpha =\Gamma _{\ \beta \gamma
}^\alpha +\widehat{\Gamma }_{\ \beta \gamma }^\alpha , 
\]
where the coefficients of deformation d--tensor $\widehat{\Gamma }_{\ \beta
\gamma }^\alpha =\{\widehat{L}_{jk}^i,\widehat{L}_{bk}^a,\widehat{C}_{jc}^i,%
\widehat{C}_{bc}^a\}$ are computed by introducing the values (\ref{conft})
into (\ref{dcon}), 
\begin{eqnarray}  \label{defcon}
\widehat{L}_{jk}^i &=& \delta _j^i\psi _k+\delta _k^i\psi
_j-g_{jk}g^{in}\psi _n, \qquad \widehat{L}_{bk}^a = \delta _b^a\psi _k, \\
\widehat{C}_{jc}^i &=& \delta _j^i\psi _c,\ \widehat{C}_{bc}^a = \delta
_b^a\psi _c+\delta _c^a\psi _b-h_{bc}h^{ae}\psi _e  \nonumber
\end{eqnarray}
with $\delta _j^i$ and $\delta _b^a$ being corresponding Kronecker symbols
in h-- and v--subspaces and 
\[
\psi _i=\delta _i\ln \Omega \mbox{ and }\psi _a=\partial _a\ln \Omega . 
\]
In this subsection we present the general formulas for a $n$--dimensional $h$%
--subspace, with indices $i,j,k...=1,2,...n,$ and $m$--dimensional $v$%
--subspace, with indices $a,b,c,...=1,2,...m.$

The d--connection deformations (\ref{defcon}) induce conformal deformations
of the Ricci d--tensor (\ref{dricci}), 
\begin{eqnarray}
\widetilde{R}_{hj} &=& R_{hj}+\widehat{R}_{[1]hj}+\widehat{R}_{[2]hj},\ 
\widetilde{R}_{ja}= R_{ja}+\widehat{R}_{ja},  \nonumber \\
\widetilde{R}_{bk} &= & R_{bk}+ \widehat{R}_{bk},\ \widetilde{S}_{bc} =
S_{bc}+\widehat{S}_{bc},  \nonumber
\end{eqnarray}
where the deformation Ricci d--tensors are 
\begin{eqnarray}
\widehat{R}_{[1]hj} &=& \partial _i\widehat{L}_{hj}^i-\partial _j\widehat{L}%
_h+\widehat{L}_{\ hj}^mL_m+L_{\ hj}^m\widehat{L}_m+ \widehat{L}_{\ hj}^m%
\widehat{L}_m -\widehat{L}_{\ hi}^mL_{mj}^i-L_{\ hi}^m\widehat{L}_{mj}^i- 
\widehat{L}_{\ hi}^m\widehat{L}_{mj}^i,  \nonumber \\
\widehat{R}_{[2]hj} &=& N_i^a\partial _a\widehat{L}_{hj}^i-N_j^a\partial _a%
\widehat{L}_h+\widehat{C}_{ha}^iR_{\ ji}^a;  \label{driccid} \\
\widehat{R}_{ja} &=& -\partial _a\widehat{L}_j+\delta _i\widehat{C}%
_{ja}^i+L_{ki}^i\widehat{C}_{ja}^k-L_{ji}^k\widehat{C}_{ka}^i-L_{ai}^b%
\widehat{C}_{jb}^i -\widehat{C}_{\ jb}^iP_{\ ia}^b-C_{\ jb}^i\widehat{P}_{\
ia}^b- \widehat{C}_{\ jb}^i\widehat{P}_{\ ia}^b,  \nonumber \\
\widehat{R}_{bk} &=& \partial _a\widehat{L}_{\ bk}^a-\delta _k\widehat{C}_b+
L_{\ bk}^a\widehat{C}_a +\widehat{C}_{\ bd}^aP_{\ ka}^d + C_{\ bd}^a\widehat{%
P}_{\ ka}^d+ \widehat{C}_{\ bd}^a\widehat{P}_{\ ka}^d,  \nonumber \\
\widehat{S}_{bc} &=& \partial _a\widehat{C}_{\ bc}^a-\partial _c\widehat{C}%
_b+ \widehat{C}_{\ bc}^eC_e+C_{\ bc}^e\widehat{C}_e+ \widehat{C}_{\ bc}^e%
\widehat{C}_e - \widehat{C}_{\ ba}^eC_{\ ec}^a - C_{\ ba}^e\widehat{C}_{\
ec}^a- \widehat{C}_{\ ba}^e\widehat{C}_{\ ec}^a,  \nonumber
\end{eqnarray}
when $\widehat{L}_h=\widehat{L}_{hi}^i$ and $\widehat{C}_b=\widehat{C}_{\
be}^e.$

\subsection{An ansatz with adapted conformal factor and N--connection}

We consider a 3D metric 
\begin{equation}  \label{ansatzc}
g_{\alpha \beta }=\left[ 
\begin{array}{ccc}
\Omega ^2(a-w_1^{\ 2}h) & -w_1w_2h\Omega ^2 & -w_1h\Omega ^2 \\ 
-w_1w_2h\Omega ^2 & \Omega ^2(b-w_2^{\ 2}h) & -w_2h\Omega ^2 \\ 
-w_1h\Omega ^2 & -w_2h\Omega ^2 & -h\Omega ^2%
\end{array}
\right]
\end{equation}
where $a=a(x^i),b=b\left( x^i\right) ,w_i=w_i(x^k,y),\Omega =\Omega \left(
x^k,y\right) \geq 0$ and $h=h\left( x^k,y\right) $ when the conditions 
\[
\psi _i=\delta _i\ln \Omega =\frac \partial {\partial x^i}\ln \Omega -w_i\ln
\Omega =0 
\]
are satisfied. With respect to anholonomic bases (\ref{ddif}) the (\ref%
{ansatzc}) transforms into the d--metric 
\begin{equation}  \label{dansatzc}
\delta s^2 = \Omega ^2(x^k,y)[a(x^k)(dx^1)^2+b(x^k)(dx^1)^2 +h\left(
x^k,y\right) (\delta y)^2].
\end{equation}

By straightforward calculus, by applying consequently the formulas (\ref%
{dcurvatures})--(\ref{einsteq2d}) we find that there is a non--trivial
coefficient of the Ricci d--tensor (\ref{dricci}), of the deformation
d--tensor (\ref{driccid}), 
\[
\widehat{R}_{j3}=\psi _3\cdot \delta _j\ln \sqrt{|h|}, 
\]
which results in non--trivial components of the Einstein d-tensor (\ref%
{einstdt}), 
\[
G_3^3=-hR_1^1 \mbox{ and } P_{\bullet i}=-\psi _3\cdot \delta _j\ln \sqrt{|h|%
}, 
\]
where $R_1^1$ is given by the formula (\ref{ricci1d}).

We can select a class of solutions of 3D Einstein equations with $P_{\bullet
j}=0$ but with the horizontal components of metric depending on anisotropic
coordinate $y,$ via conformal factor $\Omega (x^{k},y),$ and dynamical
components of the N--connection, $w_{i},$ if we choose 
\[
h(x^{k},y)=\pm \Omega ^{2}(x^{k},y) 
\]
and state 
\begin{equation}
w_{i}(x^{k},y)=\partial _{i}\ln |\ln \Omega |.  \label{conformnc}
\end{equation}
Finally, we not that for the ansatz (\ref{ansatzc}) (equivalently (\ref%
{dansatzc})) the coefficients of N--connection have to be found as dynamical
values by solving the Einstein equations.

\section{Matter Energy Momentum D--Tensors}

\subsection{Variational definition of energy-momentum d--tensors}

For locally isotropic spacetimes the symmetric energy momentum tensor is to
be computed by varying on the metric (see, for instance, Refs. \cite{haw,mtw}%
) the matter action 
\[
S=\frac 1c\int {\cal L}\sqrt{|g|}dV, 
\]
where ${\cal L}$ is the Lagrangian of matter fields, $c$ is the light
velocity and $dV$ is the infinitesimal volume, with respect to the inverse
metric $g^{\alpha \beta }$. By definition one states that the value 
\begin{equation}  \label{emtens}
\frac 12\sqrt{|g|}T_{\alpha \beta }=\frac{\partial (\sqrt{|g|}{\cal L})}{%
\partial g^{\alpha \beta }}-\frac \partial {\partial u^\tau }\frac{\partial (%
\sqrt{|g|}{\cal L})}{\partial g^{\alpha \beta }/\partial u^\tau }
\end{equation}
is the symmetric energy--momentum tensor of matter fields. With respect to
anholonomic frames (\ref{dder}) and (\ref{ddif}) there are imposed
constraints of type 
\[
g_{ib}-N_i^{\bullet }h=0 
\]
in order to obtain the block representation for d--metric (\ref{dm}). Such
constraints, as well the substitution of partial derivatives into
N--elongated, could result in nonsymmetric energy--momentum d--tensors $%
\Upsilon _{\alpha \beta }$ which is compatible with the fact that on a
locally anisotropic spacetime the Ricci d--tensor could be nonsymmetric.

The gravitational--matter field interactions on locally anisotropic
spa\-ce\-ti\-mes are described by dynamical models with imposed constraints
(a generalization of anholonomic analytic mechanics for gravitational field
theory). The physics of systems with mixed holonomic and anholonomic
variables states additional tasks connected with the definition of
conservation laws, interpretation of non--symmetric energy--momentum tensors 
$\Upsilon _{\alpha \beta}$ on locally anisotropic spacetimes and relation of
such values with, for instance, the non--symmetric Ricci d--tensor. In this
work we adopt the convention that for locally anisotropic gravitational
matter field interactions the non--symmetric Ricci d--tensor induces a
non--symmetric Einstein d--tensor which has as a source a corresponding
non--symmetric matter energy--momentum tensor. The values $\Upsilon _{\alpha
\beta}$ should be computed by a variational calculus on locally anisotropic
spacetime as well by imposing some constraints following the symmetry of
anisotropic interactions and boundary conditions.

In the next subsection we shall investigate in explicit form some cases of
definition of energy momentum tensor for locally anisotropic matter on
locally anisotropic spacetime.

\subsection{Energy--Momentum D--Tensors for Anisotropic Media}

Following DeWitt approach \cite{dw} and recent results on dynamical collapse
and hair of black holes of Husain and Brown \cite{hus}, we set up a
formalism for deriving energy--momentum d--tensors for locally anisotropic
matter.

Our basic idea for introducing a local anisotropy of matter is to rewrite
the energy--momentum tensors with respect to locally adapted frames and to
change the usual partial derivations and differentials into corresponding
operators (\ref{dder}) and (\ref{ddif}), ''elongated'' by N--connection. The
energy conditions (weak, dominant, or strong) in a locally anisotropic
background have to be analyzed with respect to a locally anisotropic basis.

We start with DeWitt's action written in locally anisotropic spacetime, 
\[
S\left[ g_{\alpha \beta },z^{\underline{i}}\right] =-\int\limits_V\delta ^3u%
\sqrt{-g}\rho \left( z^{\underline{i}},q_{\underline{j}\underline{k}}\right)
, 
\]
as a functional on region $V,$ of the locally anisotropic metric $g_{\alpha
\beta }$ and the Lagrangian coordinates $z^{\underline{i}}=z^{\underline{i}%
}\left( u^\alpha \right) $ (we use underlined indices $\underline{i},%
\underline{j},...=1,2$ in order to point out that the 2--dimensional matter
space could be different from the locally anisotropic spacetime). The
functions $z^{\underline{i}}=z^{\underline{i}}\left( u^\alpha \right) $ are
two scalar locally anisotropic fields whose locally anisotropic gradients
(with partial derivations substituted by operators (\ref{pder})) are
orthogonal to the matter world lines and label which particle passes through
the point $u^\alpha .$ The action $S\left[ g_{\alpha \beta },z^{\underline{i}%
}\right] $ is the proper volume integral of the proper energy density $\rho $
in the rest anholonomic frame of matter. The locally anisotropic density $%
\rho \left( z^{\underline{i}},q_{\underline{j}\underline{k}}\right) $
depends explicitly on $z^{\underline{i}}$ and on matter space d--metric $q^{%
\underline{i}\underline{j}}=\left( \delta _\alpha z^{\underline{i}}\right)
g^{\alpha \beta }\left( \delta _\beta z^{\underline{j}}\right) ,$ which is
interpreted as the inverse d--metric in the rest anholonomic frame of the
matter.

Using the d--metric $q^{\underline{i}\underline{j}}$ and locally anisotropic
fluid velocity $V^\alpha ,$ defined as the future pointing unit d--vector
orthogonal to d--gradients $\delta _\alpha z^{\underline{i}},$ the locally
anisotropic spacetime d--metric (\ref{dmetric}) of signature (--,+,+) may be
written in the form 
\[
g_{\alpha \beta }=-V_\alpha V_\beta +q_{\underline{j}\underline{k}}\delta
_\alpha z^{\underline{j}}\delta _\beta z^{\underline{k}} 
\]
which allow us to define the energy--momentum d--tensor for elastic locally
anisotropic medium as 
\begin{equation}  \label{emem}
\Upsilon _{\beta \gamma } \equiv -\frac 2{\sqrt{-g}}\frac{\delta S}{\delta
g^{\beta \gamma }} \rho V_\beta V_\gamma + t_{\underline{j}\underline{k}%
}\delta _\beta z^{\underline{j}} \delta _\gamma z^{\underline{k}} , 
\nonumber
\end{equation}
where the locally anisotropic matter stress d--tensor $t_{\underline{j}%
\underline{k}}$ is expressed as 
\begin{eqnarray}
t_{\underline{j}\underline{k}}= 2\frac{\delta \rho } {\partial q^{\underline{%
j}\underline{k}}}- \rho q_{\underline{j}\underline{k}}= \frac 2{\sqrt{q}} 
\frac{\delta \left( \sqrt{q}\rho \right) } {\partial q^{\underline{j}%
\underline{k}}}.  \label{dem1}
\end{eqnarray}

Here one should be noted that on locally anisotropic spaces 
\[
D_\alpha \Upsilon ^{\alpha \beta }=D_\alpha \left( R^{\alpha \beta }-\frac
12g^{\alpha \beta }R\right) =J^\beta \neq 0 
\]
and this expression must be treated as a generalized type of conservation
law with a geometric source $J^\beta $ for the divergence of locally
anisotropic matter d--tensor \cite{ma}.

The stress--energy--momentum d--tensor for locally an\-iso\-trop\-ic elastic
medium is defined by applying N--elongated operators $\delta _\alpha $ of
partial derivatives (\ref{pder}), 
\begin{equation}
T_{\alpha \beta } = -\frac 2{\sqrt{-g}}\frac{\delta S}{\delta g^{\alpha
\beta }} = -\rho g_{\alpha \beta }+ 2\frac{\partial \rho }{\partial q^{%
\underline{i} \underline{j}}}\delta _\alpha z^{\underline{j}} \delta _\beta
z^{\underline{k}} = -V_\alpha V_\beta +\tau _{^{\underline{i} \underline{j}%
}}\delta _\alpha z^{\underline{i}} \delta _\beta z^{\underline{j}}, 
\nonumber
\end{equation}
where we introduce the matter stress d--tensor 
\[
\tau _{^{\underline{i}\underline{j}}}=2\frac{\partial \rho }{\partial q^{%
\underline{i}\underline{j}}}-\rho q_{\underline{i}\underline{j}}=\frac 2{%
\sqrt{q}}\frac{\partial \left( \sqrt{q}\rho \right) }{\partial q^{\underline{%
i}\underline{j}}}. 
\]
The obtained formulas generalize for spaces with nontrivial N--connection
structures the results on isotropic and anisotropic media on locally
isotropic spacetimes.

\subsection{Isotropic and anisotropic media}

The {\it isotropic elastic}, but in general locally anisotropic medium is
introduced as one having equal all principal pressures with stress d--tensor
being for a perfect fluid and the density $\rho =\rho \left( n\right) ,$
where the proper density (the number of particles per unit proper volume in
the material rest anholonomic frame) is $n=\underline{n}\left( z^{\underline{%
i}}\right) /\sqrt{q};$ the value $\underline{n}\left( z^{\underline{i}%
}\right) $ is the number of particles per unit coordinate cell $\delta ^3z.$
With respect to a locally anisotropic frame, using the identity 
\[
\frac{\partial \rho \left( n\right) }{\partial q^{\underline{j}\underline{k}}%
}=\frac n2\frac{\partial \rho }{\partial n}q_{\underline{j}\underline{k}} 
\]
in (\ref{dem1}), the energy--momentum d--tensor (\ref{emem}) 
for a isotropic elastic locally anisotropic medium becomes 
\[
\Upsilon _{\beta \gamma }=\rho V_\beta V_\gamma +\left( n\frac{\partial \rho 
}{\partial n}-\rho \right) \left( g_{\beta \gamma }+V_\beta V_\gamma \right)
. 
\]
This medium looks like isotropic with respect to anholonomic frames but, in
general, it is locally anisotropic.

The {\it anisotropic elastic} and locally anisotropic medium has not equal
principal pressures. In this case we have to introduce (1+1) decompositions
of locally anisotropic matter d--tensor $q_{\underline{j}\underline{k}}$%
\[
q_{\underline{j}\underline{k}}=\left( 
\begin{array}{cc}
\alpha ^2+\beta ^2 & \beta \\ 
\beta & \sigma%
\end{array}
\right) , 
\]
and consider densities $\rho \left( n_{\underline{1}},n_{\underline{2}%
}\right) ,$ where $n_{\underline{1}}$ and $n_{\underline{2}}$ are
respectively the particle numbers per unit length in the directions given by
bi--vectors $v_{\underline{j}}^{\underline{1}}$ and $v_{\underline{j}}^{%
\underline{2}}.$ Substituting 
\[
\frac{\partial \rho \left( n_{\underline{1}},n_{\underline{2}}\right) }{%
\partial h^{\underline{j}\underline{k}}}=\frac{n_{\underline{1}}}2\frac{%
\partial \rho }{\partial n_{\underline{1}}}v_{\underline{j}}^{\underline{1}%
}v_{\underline{k}}^{\underline{1}}+\frac{n_{\underline{2}}}2\frac{\partial
\rho }{\partial n_{\underline{2}}}v_{\underline{j}}^{\underline{2}}v_{%
\underline{k}}^{\underline{2}} 
\]
into (\ref{dem1}), 
which gives 
\[
t_{\underline{j}\underline{k}}=\left( n_{\underline{1}}\frac{\partial \rho }{%
\partial n_{\underline{1}}}-\rho \right) v_{\underline{j}}^{\underline{1}}v_{%
\underline{k}}^{\underline{1}}+\left( n_{\underline{2}}\frac{\partial \rho }{%
\partial n_{\underline{2}}}-\rho \right) v_{\underline{j}}^{\underline{2}}v_{%
\underline{k}}^{\underline{2}}, 
\]
we obtain from (\ref{emem}) 
the energy--momentum d--tensor for the anisotropic locally anisotropic
matter 
\begin{equation}
\Upsilon _{\beta \gamma } = \rho V_\beta V_\gamma + \left( n_{\underline{1}}%
\frac{\partial \rho }{\partial n_{\underline{1}}}- \rho \right) v_{%
\underline{j}}^{\underline{1}} v_{\underline{k}}^{\underline{1}}+\left( n_{%
\underline{2}}\frac{\partial \rho }{\partial n_{\underline{2}}}- \rho
\right) v_{\underline{j}}^{\underline{2}} v_{\underline{k}}^{\underline{2}}.
\nonumber
\end{equation}
So, the pressure $P_1=\left( n_{\underline{1}}\frac{\partial \rho }{\partial
n_{\underline{1}}}-\rho \right) $ in the direction $v_{\underline{j}}^{%
\underline{1}}$ differs from the pressure\newline
$P_2=\left( n_{\underline{2}}\frac{\partial \rho }{\partial n_{\underline{2}}%
}-\rho \right) $ in the direction $v_{\underline{j}}^{\underline{2}}.$ For
instance, if for the (2+1)--dimensional locally anisotropic spacetime we
impose the conditions $\Upsilon _1^1=\Upsilon _2^2\neq \Upsilon _3^3,$ when 
\[
\rho =\rho \left( n_{\underline{1}}\right) ,z^{\underline{1}}\left( u^\alpha
\right) =r,z^{\underline{2}}\left( u^\alpha \right) =\theta , 
\]
$r$ and $\theta $ are correspondingly radial and angle coordinates on
locally anisotropic spacetime, we have 
\begin{equation}
\Upsilon _1^1=\Upsilon _2^2=\rho ,\Upsilon _3^3=\left( n_{\underline{1}}%
\frac{\partial \rho }{\partial n_{\underline{1}}}-\rho \right) .
\end{equation}

We shall also consider the variant when the coordinated $\theta $ is
anisotropic $(t$ and $r$ being isotropic). In this case we shall impose the
conditions $\Upsilon _1^1\neq \Upsilon _2^2=\Upsilon _3^3$ for 
\[
\rho =\rho \left( n_{\underline{1}}\right) ,z^{\underline{1}}\left( u^\alpha
\right) =t,z^{\underline{2}}\left( u^\alpha \right) =r 
\]
and 
\begin{equation}  \label{dtema}
\Upsilon _1^1=\left( n_{\underline{1}}\frac{\partial \rho }{\partial n_{%
\underline{1}}}-\rho \right) ,\Upsilon _2^2=\Upsilon _3^3=\rho ,.
\end{equation}

The anisotropic elastic locally anisotropic medium described here satisfies
respectively weak, dominant, or strong energy conditions only if the
corresponding restrictions are placed on the equation of state considered
with respect to an anholonomic frame (see Ref. \cite{hus} for similar
details in locally isotropic cases). For example, the weak energy condition
is characterized by the inequalities $\rho \geq 0$ and $\partial \rho
/\partial n_{\underline{1}}\geq 0.$

\subsection{Spherical symmetry with respect to holonomic and anholonomic
frames}

In radial coordinates $\left( t,r,\theta \right) $ (with $-\infty \leq
t<\infty ,$ $0\leq r<\infty ,$ $0\leq \theta \leq 2\pi )$ for a spherically
symmetric 3D metric (\ref{m2}) 
\begin{equation}  \label{m3}
ds^2=-f\left( r\right) dt^2+\frac 1{f\left( r\right) }dr^2+r^2d\theta ^2,
\end{equation}
with the energy--momentum tensor (\ref{emtens}) written 
\begin{equation}
T_{\alpha \beta } = \rho \left( r\right) \left( v_\alpha w_\beta + v_\beta
w_\alpha \right) + P\left( r\right) \left( g_{\alpha \beta} + v_\alpha
w_\beta +v_\beta w_\alpha \right) ,  \nonumber
\end{equation}
where the null d--vectors $v_\alpha $ and $w_\beta $ are defined by 
\begin{eqnarray}
V_\alpha & = & \left( \sqrt{f},-\frac 1{\sqrt{f}},0\right) =\frac 1{\sqrt{2}%
}\left( v_\alpha +w_\alpha \right) ,  \nonumber \\
q_\alpha & = & \left( 0,\frac 1{\sqrt{f}},0\right) =\frac 1{\sqrt{2}}\left(
v_\alpha -w_\alpha \right) .  \nonumber
\end{eqnarray}

In order to investigate the dynamical spherically symmetric \cite{cm}
collapse solutions it is more convenient to use the coordinates $\left(
v,r,\theta \right) ,$ where the advanced time coordinate $v$ is defined by $%
dv=dt+\left( 1/f\right) dr.$ The metric (\ref{m3}) may be written 
\begin{equation}  \label{m4}
ds^2=-e^{2\psi \left( v,r\right) }F\left( v,r\right) dv^2+2e^{\psi \left(
v,r\right) }dvdr+r^2\theta ^2,
\end{equation}
where the mass function $m\left( v,r\right) $ is defined by $F\left(
v,r\right) =1-2m\left( v,r\right) /r.$ Usually, one considers the case $\psi
\left( v,r\right) =0$ for the type II \cite{haw} energy--momentum d--tensor 
\begin{equation}
T_{\alpha \beta } = \frac 1{2\pi r^2}\frac{\delta m}{\partial v}v_\alpha
v_\beta + \rho \left( v,r\right) \left( v_\alpha w_\beta +v_\beta w_\alpha
\right) + P\left( v,r\right) \left( g_{\alpha \beta }+ v_\alpha w_\beta +
v_\beta w_\alpha \right)  \nonumber
\end{equation}
with the eigen d-vectors $v_\alpha =\left( 1,0,0\right) $ and $w_\alpha
=\left( F/2,-1,0\right) $ and the non--vanishing components 
\begin{eqnarray}
T_{vv} & = & \rho \left( v,r\right) \left( 1-\frac{2m\left( v,r\right) }%
r\right) +\frac 1{2\pi r^2}\frac{\delta m\left( v,r\right) } {\partial v},
\label{dem4b} \\
T_{vr} & = & -\rho \left( v,r\right) ,\quad T_{\theta \theta }=P\left(
v,r\right) g_{\theta \theta }.  \nonumber
\end{eqnarray}

To describe a locally isotropic collapsing pulse of radiation one may use
the metric 
\begin{equation}  \label{m4a}
d s^2 = \left[ \Lambda r^2+m\left( v\right) \right] dv^2 + 2dvdr - j\left(
v\right) dvd\theta +r^2 d \theta ^2,
\end{equation}
with the Einstein field equations (\ref{einst1}) reduced to 
\[
\frac{\partial m\left( v\right) }{dv}=2\pi \rho \left( v\right) ,\frac{%
\partial j\left( v\right) }{dv}=2\pi \omega \left( v\right) 
\]
having non--vanishing com\-po\-nents of the energy--mo\-mentum d--tensor
(for a rotating null locally anisotropic fluid), 
\begin{equation}  \label{dem4a}
T_{vv}=\frac{\rho \left( v\right) }r+\frac{j\left( v\right) \omega \left(
v\right) }{2r^3},\quad T_{v\theta }=-\frac{\omega \left( v\right) }r,
\end{equation}
where $\rho \left( v\right) $ and $\omega \left( v\right) $ are arbitrary
functions.

In a similar manner we can define energy--momentum d--tensors for various
systems of locally anisotropic distributed matter fields; all values have to
be re--defined with respect to anholonomic bases of type (\ref{dder}) and (%
\ref{ddif}). For instance, let us consider the angle $\theta $ as the
anisotropic variable. In this case we have to 'elongate' the differentials, 
\[
d\theta \rightarrow \delta \theta =d\theta +w_i\left( v,r,\theta \right)
dx^i, 
\]
for the metric (\ref{m4}) (or (\ref{m4a})), by transforming it into a
d--metric, substitute all partial derivatives into N--elongated ones, 
\[
\partial _i\rightarrow \delta _i=\partial _i-w_i\left( v,r,\theta \right)
\frac \partial {\partial \theta }, 
\]
and 'N--extend' the operators defining the Riemanni\-an, Ricci, Einstein and
energy--mo\-ment\-um tensors $T_{\alpha \beta },$ transforming them into
respective d--tensors. We compute the components of the energy--momentum
d--tensor for elastic media as the coefficients of usual energy--momentum
tensor redefined with respect to locally anisotropic frames, 
\begin{eqnarray}
\Upsilon _{11} &=& T_{11}+\left( w_1\right) ^2T_{33}, \Upsilon _{33}=T_{33}
\label{demelm} \\
\Upsilon _{22} &=& T_{22}+ \left( w_2\right) ^2 T_{33}, \Upsilon _{12} =
\Upsilon _{21}=T_{21}+w_2w_1T_{33},  \nonumber \\
\Upsilon _{i3}&=& T_{i3}+w_iT_{33},\Upsilon _{3i}=T_{3i}+w_iT_{33}, 
\nonumber
\end{eqnarray}
where the $T_{\alpha \beta }$ are given by the coefficients (\ref{dem4b})
(or (\ref{dem4a})). If the isotropic energy--momentum tensor does not
contain partial derivatives, the corresponding d--tensor is also symmetric
which is less correlated with the possible antisymmetry of the Ricci tensor
(for such configurations we shall search solutions with vanishing
antisymmetric components).

\section{3D Solutions Induced by Anisotropic Matter}

We investigate a new class of solutions of (2+1)--dimensional Einstein
equations coupled with anisotropic matter \cite{cm,hus,btz,chan,ross} which
describe locally anisotropic collapsing configurations.

Let us consider the locally isotropic metric 
\begin{equation}  \label{m5}
\widehat{g}_{\alpha \beta }=\left[ 
\begin{array}{ccc}
g\left( v,r\right) & 1/2 & 0 \\ 
1/2 & 0 & 0 \\ 
0 & 0 & r^2%
\end{array}
\right]
\end{equation}
which solves the locally isotopic variant of Einstein equations (\ref{einst1}%
) if 
\begin{equation}  \label{m5a}
g\left( v,r\right) =-[1-2g(v)-2h(v)r^{1-k}-\Lambda r^2],
\end{equation}
where the functions $g(v)$ and $h(v)$ define the mass function 
\[
m(r,v)=g(v)r+h(v)r^{2-k}+\frac \Lambda 2r^2 
\]
satisfying the dominant energy conditions 
\[
P\geq 0,\rho \geq P,T_{ab}w^aw^b>0 
\]
if 
\[
\frac{dm}{dv}=\frac{dg}{dv}r+\frac{dh}{dv}r^{2-k}>0. 
\]
Such solutions of the Vadya type with locally isotropic null fluids have
been considered in Ref. \cite{hus}.

\subsection{Solutions with generic local anisotropy in spherical coordinates}

By introducing a new time--like variable 
\[
t=v+\int \frac{dr}{g\left( v,r\right) } 
\]
the metric(\ref{m5}) can be transformed in diagonal form 
\begin{equation}  \label{m6}
ds^2=-g\left( t,r\right) dt^2+\frac 1{g\left( t,r\right) }dr^2+r^2d\theta ^2
\end{equation}
which describe the locally isotropic collapse of null fluid matter.

A variant of locally anisotropic inhomogeneous collapse could be modeled,
for instance, by the N--elongation of the variable $\theta $ in (\ref{m6})
and considering solutions of vacuum Einstein equations for the d--metric (a
particular case of (\ref{dm3})) 
\begin{equation}
ds^{2}=-g\left( t,r\right) dt^{2}+\frac{1}{g\left( t,r\right) }%
dr^{2}+r^{2}\delta \theta ^{2},  \label{dm3}
\end{equation}
where 
\[
\delta \theta =d\theta +w_{1}\left( t,r,\theta \right) dt+w_{2}\left(
t,r,\theta \right) dr. 
\]
The coefficients $g\left( t,r\right) ,1/g\left( t,r\right) $ and $r^{2}$ of
the d--metric were chosen with the aim that in the locally isotropic limit,
when $w_{i}\rightarrow 0,$ we shall obtain the 3D metric (\ref{m6}). We note
that the gravitational degrees of freedom are contained in nonvanishing
values of the Ricci d--tensor (\ref{dricci}), 
\begin{equation}
R_{1}^{1}=R_{2}^{2}=\frac{1}{2g^{3}}[(\frac{\partial g}{\partial r}%
)^{2}-g^{3}\frac{\partial ^{2}g}{\partial t^{2}}-g\frac{\partial ^{2}g}{%
\partial r^{2}}],  \label{ricci2}
\end{equation}
of the N--curvature (\ref{ncurv}), 
\[
\Omega _{12}^{3}=-\Omega _{21}^{3}=\frac{\partial w_{1}}{\partial r}-\frac{%
\partial w_{2}}{\partial t}-w_{2}\frac{\partial w_{1}}{\partial \theta }%
+w_{1}\frac{\partial w_{2}}{\partial \theta }, 
\]
and d--torsion (\ref{dtorsions}) 
\[
P_{13}^{3}=\frac{1}{2}\left( 1+r^{4}\right) \frac{\partial w_{1}}{\partial
\theta },\ P_{23}^{3}=\frac{1}{2}\left( 1+r^{4}\right) \frac{\partial w_{2}}{%
\partial \theta }-r^{3}. 
\]

We can construct a solution of 3D Einstein equations with cosmological
constant $\Lambda $ (\ref{einst1}) and energy momentum d--tensor $\Upsilon
_{\alpha \beta },$ when $\Upsilon _{ij}=T_{ij}+w_iw_jT_{33},\Upsilon
_{3j}=T_{3j}+w_jT_{33}$ and $\Upsilon _{33}=T_{33}$ when $T_{\alpha \beta }$
is given by a d--tensor of type (\ref{dtema}), $T_{\alpha \beta }=\{n_{%
\underline{1}}\frac{\partial \rho }{\partial n_{\underline{1}}},P,0\}$ with
anisotropic matter pressure $P.$ A self--consistent solution is given by 
\begin{equation}  \label{auxw1}
\Lambda = \kappa n_{\underline{1}}\frac{\partial \rho } {\partial n_{%
\underline{1}}}=\kappa P, \mbox{ and } h=\frac{\kappa \rho }{R_1^1+\Lambda }
\end{equation}
where $R_1^1$ is computed by the formula (\ref{ricci2}) for arbitrary values 
$g\left( t,r\right) .$ For instance, we can take the $g(\nu ,r)$ from (\ref%
{m5a}) with $\nu \rightarrow t=\nu +\int g^{-1}(\nu ,r)dr.$

For $h=r^2,$ the relation (\ref{auxw1}) results in an equation for $g(t,r),$ 
\[
(\frac{\partial g}{\partial r})^2-g^3\frac{\partial ^2g}{\partial t^2}-g%
\frac{\partial ^2g}{\partial r^2}=2g^3\left( \frac{\kappa \rho }{r^2}%
-\Lambda \right) . 
\]
The static configurations are described by the equation 
\begin{equation}  \label{auxw2}
gg^{\prime \prime }-(g^{\prime })^2+\varpi (r)g^3=0,
\end{equation}
where 
\[
\varpi (r)=2\left( \frac{\kappa \rho \left( r\right) }{r^2}-\Lambda \right) 
\]
and the prime denote the partial derivative $\partial /\partial r.$ There
are four classes (see (\cite{kamke})) of solutions of the equation (\ref%
{auxw2}), which depends on constants of the relation 
\[
(\ln |g|)^{\prime }=\pm \sqrt{2|\varpi (r)|(C_1\mp g)}, 
\]
where the minus (plus) sign under square root is taken for $\varpi (r)>0$ $%
(\varpi (r)<0)$ and the constant $C_1$ can be negative, $C_1=-c^2,$ or
positive, $C_1=c^2.$ In explicit form the solutions are 
\begin{equation}  \label{faze}
g(r)=\left\{ 
\begin{array}{rcl}
c^{-22}\cosh ^{-2}\left[ \frac c2\sqrt{2|\varpi (r)|}\left( r-C_2\right) %
\right] & , &  \\ 
\mbox{ for } \varpi (r)>0,C_1=c^2 & ; &  \\ 
c^{-2}\sinh ^{-2}\left[ \frac c2\sqrt{2|\varpi (r)|}\left( r-C_2\right) %
\right] & , &  \\ 
\mbox{ for } \varpi (r)<0,C_1=c^2 & ; &  \\ 
c^{-2}\sin ^{-2}\left[ \frac c2\sqrt{2|\varpi (r)|}\left( r-C_2\right) %
\right] \neq 0 & , &  \\ 
\mbox{ for } \varpi (r)<0,C_1=-c^2 & ; &  \\ 
-2{\varpi (r)}^{-1}(r-C_2)^{-2}{\qquad}{\qquad} & , &  \\ 
\mbox{ for } \varpi (r)<0,C_1=0, &  & 
\end{array}
\right.
\end{equation}
where $C_2=const.$ The values of constants are to be found from boundary
conditions. In dependence of prescribed type of matter density distribution
and of values of cosmological constant one could fix one of the four classes
of obtained solutions with generic local anisotropy of 3D Einstein equations.

The constructed in this section static solutions of 3D Einstein equations
are locally anisotropic alternatives (with proper phases of anisotropic
polarizations of gravitational field) to the well know BTZ solution. Such
configurations are possible if anholonomic frames with associated
N--connection structures are introduced into consideration.

\subsection{An anisotropic solution in ($\protect\nu ,r,\protect\theta )$%
--coordinates}

For modeling a spherical collapse with generic local anisotropy we use the
d--metric (\ref{dm2a}) by stating the coordinates $x^1=v,x^2=r$ and $%
y=\theta .$ The equations (\ref{einst1}) are solved if 
\[
\kappa \rho (v,r)=\Lambda \mbox{ and } \kappa P(v,r)=-\Lambda -\frac 12\frac{%
\partial ^2g}{\partial r^2} 
\]
for 
\[
g=\frac \kappa \Lambda \left[ \rho (v,r)\left( 1-\frac{2m(v,r)}r\right)
+\frac 1{2\pi r^2}\frac{\delta m(v,r)}{\delta v}\right] . 
\]

\subsection{A solution for rotating two locally anisotropic fluids}

The anisotropic configuration from the previous subsection admits a
generalization to a two fluid elastic media, one of the fluids being of
locally anisotropic rotating configuration. For this model we consider an
anisotropic extension of the metric (\ref{m4a}) and of the sum of
energy--momentum tensors (\ref{dem4b}) and (\ref{dem4a}). The coordinates
are parametrized $x^1=v,x^2=r,y=\theta $ and the d--metric is given by the
ansatz 
\[
g_{ij}=\left( 
\begin{array}{cc}
g(v,r) & 1 \\ 
1 & 0%
\end{array}
\right) \mbox{ and }h=h(v,r,\theta ). 
\]
The nontrivial componenti of the Einstein d--tensor is 
\[
G_{33}=-\frac 12h\frac{\partial ^2g}{\partial r^2}. 
\]
We consider a non--rotating fluid component with nontrivial energy--momentum
components 
\begin{equation}  \label{dem5b}
{}^{(1)}T_{vv} = {}^{(1)} \rho \left( v,r\right) \left( 1-\frac{2
{}^{(1)}m\left( v,r\right) }r\right) + \frac 1{2\pi r^2}\frac{\delta
{}^{(1)} m\left( v,r\right) }{\partial v}, \ {}^{(1)}T_{vr} = - {}^{(1)}\rho
\left( v,r\right).  \nonumber
\end{equation}
and a rotating null locally anisotropic fluid with energy--momentum
components 
\begin{equation}  \label{dem5a}
^{(2)}T_{vv} = \frac{^{(2)}\rho \left( v\right) }r + \frac{%
^{(2)}j\left(v\right) \ ^{(2)}\omega \left( v\right) }{2r^3}, \
^{(2)}T_{v\theta } = -\frac{^{(2)}\omega \left( v\right) }r.  \nonumber
\end{equation}

The nontrivial components of energy momentum d--tensor $\Upsilon _{\alpha
\beta} = {}^{(1)} \Upsilon _{\alpha \beta} + {}^{(2)} \Upsilon _{\alpha
\beta}$ (associated in the locally anisotropic limit to (\ref{dem4a}) and/or
(\ref{dem4b})) are computed by using the formulas (\ref{dem5a}), (\ref{dem5b}%
) and (\ref{demelm}).

The Einstein equations are solved by the set of func\-ti\-ons 
\[
g(v,r), {}^{(1)}\rho \left( v,r\right), {}^{(1)}m\left( v,r\right),
{}^{(2)}\rho \left( v\right), {}^{(2)}j\left( v\right), {}^{(2)}\omega
\left( v\right) 
\]
satisfying the conditions 
\[
g(v,r) = \frac \kappa \Lambda \left[ ^{(1)}T_{vv}+\ ^{(2)}T_{vv}\right] , %
\mbox{ and } \Lambda = \frac 12\frac{\partial ^2g}{\partial r^2}=\kappa \
^{(1)}T_{vr}, 
\]
where $h\left( v,r,\theta \right) $ is an arbitrary function which results
in nontrivial solutions for the N--connection coefficients $w_i\left(
v,r,\theta \right) $ if $\Lambda \neq 0.$ In the locally isotropic limit,
for $^{(1)}\rho ,^{(1)}m=0,$ we could take $g(v,r)=g_1(v)+\Lambda
r^2,w_1=-j(v)/(2r^2)$ and $w_2=0$ which results in a solution of the Vadya
type with locally isotropic null fluids \cite{chan}.

The main conclusion of this subsection is that we can model the 3D collapse
of inhomogeneous null fluid by using vacuum locally anisotropic
configurations polarized by an anholonomic frame in a manner as to reproduce
in the locally isotropic limit the usual BTZ geometry.

We end this section with the remark that the locally isotropic collapse of
dust without pressure was analyzed in details in Ref. \cite{ross}.

\section{Gravitational Anisotropic Polarization and Black Holes}

If we introduce in consideration anholonomic frames, locally anisotropic
black hole configurations are possible even for vacuum locally anisotropic
spacetimes without matter. Such solutions could have horizons with deformed
circular symmetries (for instance, elliptic one) and a number of unusual
properties comparing with locally isotropic black hole solutions. In this
Section we shall analyze two classes of such solutions. Then we shall
consider the possibility to introduces matter sources and analyze such
configurations of matter energy density distribution when the gravitational
locally anisotropic polarization results into constant renormalization of
constants of BTZ solution.

\subsection{Non--rotating black holes with ellipsoidal horizon}

We consider a metric (\ref{ansatzc}) for local coordinates $%
(x^{1}=r,x^{2}=\theta ,y=t),$ where $t$ is the time--like coordinate and the
coefficients are parametrized 
\begin{equation}
a(x^{i})=a\left( r\right) ,b(x^{i})=b(r,\theta )  \label{elips1}
\end{equation}
and 
\begin{equation}
h(x^{i},y)=h\left( r,\theta \right) .  \label{elips1a}
\end{equation}
The functions $a(r)$ and $\ b\left( r,\theta \right) $ and the coefficients
of nonlinear connection $w_{i}(r,\theta ,t)$ will be found as to satisfy the
vacuum Einstein equations (\ref{vaceq}) with arbitrary function $h(x^{i},y)$
(\ref{elips1a}) stated in the form in order to have compatibility with the
BTZ solution in the locally isotropic limit.

We consider a particular case of d-metrics (\ref{dansatzc}) with
coefficients like (\ref{elips1}) and (\ref{elips1a}) when 
\begin{equation}
h(r,\theta )=4\Lambda ^{3}(\theta )\left( 1-\frac{r_{+}^{2}(\theta )}{r^{2}}%
\right) ^{3}  \label{hel3}
\end{equation}
where, for instance, 
\begin{equation}
r_{+}^{2}(\theta )=\frac{p^{2}}{\left[ 1+\varepsilon \cos \theta \right] ^{2}%
}  \label{elipsrad}
\end{equation}
is taken as to construct a 3D solution of vacuum Einstein equations with
generic local anisotropy having the horizon given by the parametric equation 
\[
r^{2}=r_{+}^{2}(\theta ) 
\]
describing a ellipse with parameter $p$ and eccentricity $\varepsilon .$ We
have\ to identify 
\[
p^{2}=r_{+[0]}^{2}=-M_{0}/\Lambda _{0}, 
\]
where\ $r_{+[0]},M_{0}$ and $\Lambda _{0}$ are respectively the horizon
radius, mass parameter and cosmological constant of the non--rotating BTZ
solution \cite{btz} if we wont to have a connection with locally isotropic
limit with $\varepsilon \rightarrow 0.$ We can consider that the elliptic
horizon (\ref{elipsrad}) is modeled by the anisotropic mass 
\[
M\left( \theta \right) =M_{0}/\left[ 1+\varepsilon \cos \theta \right] ^{2}. 
\]

For the coefficients (\ref{elips1}) the equations (\ref{vaceq}) simplifies
into 
\begin{equation}  \label{vaceqel1}
-\ddot b+\frac 1{2b}\dot b^2+\frac 1{2a}\dot a\dot b=0,
\end{equation}
where (in this subsection) $\dot b=\partial b/\partial r.$ The general
solution of (\ref{vaceqel1}), for a given function $a(r)$ is defined by two
arbitrary functions $b_{[0]}(\theta )$ and $b_{[1]}(\theta )$ (see \cite%
{kamke}), 
\[
b(r,\theta )=\left[ b_{[0]}(\theta )+b_{[1]}(\theta )\int \sqrt{|a(r)|}dr%
\right] ^2. 
\]

If we identify 
\[
b_{[0]}(\theta )=2\frac{\Lambda (\theta )}{\sqrt{|\Lambda _{0}|}}%
r_{+}^{2}(\theta )\mbox{ and }b_{[1]}(\theta )=-2\frac{\Lambda (\theta )}{%
\Lambda _{0}}, 
\]
we construct a d--metric locally anisotropic solution of vacuum Einstein
equations 
\begin{equation}
\delta s^{2}=\Omega ^{2}\left( r,\theta \right) \left[ 4r^{2}|\Lambda
_{0}|dr^{2}+\frac{4}{|\Lambda _{0}|}\Lambda ^{2}(\theta )\left[
r_{+}^{2}(\theta )-r^{2}\right] ^{2}d\theta ^{2}-\frac{4}{|\Lambda _{0}|r^{2}%
}\Lambda ^{3}(\theta )\left[ r_{+}^{2}(\theta )-r^{2}\right] ^{3}\delta t^{2}%
\right] ,  \label{elipbh}
\end{equation}
where 
\[
\delta t=dt+w_{1}(r,\theta )dr+w_{2}(r,\theta )d\theta 
\]
is to be associated to a N--connection structure 
\[
w_{r}=\partial _{r}\ln |\ln \Omega |\mbox{ and }w_{\theta }=\partial
_{\theta }\ln |\ln \Omega | 
\]
with $\Omega ^{2}=\pm h(r,\theta ),$ where $h(r,\theta )$ is taken from (\ref%
{hel3}). In the simplest case we can consider a constant effective
cosmological constant $\Lambda (\theta )\simeq \Lambda _{0}.$

The matrix 
\[
g_{\alpha \beta }=\Omega ^{2}\left[ 
\begin{array}{ccc}
a-w_{1}^{\ 2}h & -w_{1}w_{2}h & -w_{1}h \\ 
-w_{1}w_{2}h & b-w_{2}^{\ 2}h & -w_{2}h \\ 
-w_{1}h & -w_{2}h & -h%
\end{array}
\right] . 
\]
parametrizes a class of solutions of 3D vacuum Einstein equations with
generic local anisotro\-py and nontrivial N--connection curvature (\ref%
{ncurv}), which describes black holes with variable mass parameter $M\left(
\theta \right) $ and elliptic horizon. As a matter of principle, by fixing
necessary functions $b_{[0]}(\theta )$ and $b_{[1]}(\theta )$ we can
construct solutions with effective (polarized by the vacuum anisotropic
gravitational field) variable cosmological constant $\Lambda (\theta ).$ We
emphasize that this type of anisotropic black hole solutions have been
constructed by solving the vacuum Einstein equations without cosmological
constant. Such type of constants or varying on $\theta $ parameters were
introduced as some values characterizing anisotropic polarizations of vacuum
gravitational field and this approach can be developed if we are considering
anholonomic frames on (pseudo) Riemannian spaces. For the examined
anisotropic model the cosmological constant is induced effectively in
locally isotropic limit via specific gravitational field vacuum
polarizations.

\subsection{Rotating black holes with running in time constants}

A new class of solutions of vacuum Einstein equations is generated by a
d--metric (\ref{dm2}) written for local coordinates $(x^1=r,x^2=t,y=\theta
), $ where as the anisotropic coordinate is considered the angle variable $%
\theta $ and the coefficients are parametrized 
\begin{equation}  \label{elips4}
a(x^i)=a\left( r\right) ,b(x^i)=b(r,t)
\end{equation}
and 
\begin{equation}  \label{elips4a}
h(x^i,y)=h\left( r,t\right) .
\end{equation}

Let us consider a 3D metric 
\begin{equation}  \label{metrbtzc}
ds^2 = 4\frac{\psi ^2}{r^2}dr^2-\frac{N_{[s]}^4\psi ^4}{r^4}dt^2 +\frac{%
N_{(s)}^2\psi ^6}{r^4}\left[ d\theta +N_{[\theta ]}dt\right] ^2  \nonumber
\end{equation}
which is conformally equivalent (if multiplied to the conformal factor $%
4N_{(s)}^2\psi ^4/r^4)$ to the rotating BTZ solution with 
\begin{eqnarray}  \label{auxform1}
N_{[s]}^2(r) &=& -\Lambda _0\frac{r^2}{\psi ^2}\left( r^2-r_{+[0]}^2\right)
,N_{[\theta ]}(r)=-\frac{J_0}{2\psi },  \nonumber \\
\psi ^2(r) &=& r^2-\frac 12\left( \frac{M_0}{\Lambda _0}+r_{+[0]}^2\right), 
\nonumber \\
r_{+[0]}^2 &=& -\frac{M_0}{\Lambda _0}\sqrt{1+\Lambda _0\left( \frac{J_0}{M_0%
}\right) ^2},  \nonumber
\end{eqnarray}
where $J_0$ is the rotation moment and $\Lambda _0$ and $M_0$ are
respectively the cosmological and mass BTZ constants.

A d--metric (\ref{dm2}) defines a locally anisotropic extension of (\ref%
{metrbtzc}) if the solution of (\ref{vaceqel1}), in variables $%
(x^1=r,x^2=t), $ with coefficients (\ref{elips4}) and (\ref{elips4a}), is
written 
\begin{equation}
b(r,t) = - \left[ b_{[0]}(t)+b_{[1]}(t)\int \sqrt{\left| a(r)\right| }dr%
\right]^2 = -\Lambda ^2(t)\left[ r_{+}^2(t)-r^2\right] ^2,  \nonumber
\end{equation}
for 
\[
a(r)=4\Lambda _0r^2,b_{[0]}(t)=\Lambda (t)r_{+}^2(t),b_{[1]}(t)=2\Lambda (t)/%
\sqrt{|\Lambda _0|} 
\]
with $\Lambda (t)\sim \Lambda _0$ and $r_{+}(t)\sim r_{+[0]}$ being some
running in time values.

The functions $a(r)$ and $\ b\left( r,t\right) $ and the coefficients of
nonlinear connection $w_i(r,t,\theta )$ must solve the vacuum Einstein
equations (\ref{vaceq}) with arbitrary function $h(x^i,y)$ (\ref{elips1a})
stated in the form in order to have a relation with the BTZ solution for
rotating black holes in the locally isotropic limit. This is possible if we
choose 
\begin{equation}
w_1(r,t) = -\frac{J\left( t\right) }{2\psi (r,t)}, \qquad h(r,t) = \frac{%
4N_{[s]}^2(r,t)\psi ^6(r,t)}{r^4},  \nonumber
\end{equation}
for an arbitrary function $w_2(r,t,\theta )$ with $N_{[s]}(r,t)$ and $\psi
(r,t)$ computed by the same formulas (\ref{auxform1}) with the constant
substituted into running values, 
\[
\Lambda _0\rightarrow \Lambda (t),M_0\rightarrow M\left( t\right)
,J_0\rightarrow J(t). 
\]

We can model a dissipation of 3D black holes, by anisotropic gravitational
vacuum polarizations if for instance, 
\[
r_{+}^2(t)\simeq r_{+[0]}^2\exp [-\lambda t] 
\]
for $M(t)=M_0\exp [-\lambda t]$ with $M_0$ and $\lambda $ being some
constants defined from some ''experimental'' data or a quantum model for 3D
gravity. The gravitational vacuum admits also polarizations with exponential
and/or oscillations in time for $\Lambda (t)$ and/or of $M(t).$

\subsection{Anisotropic Renormalization of Constants}

The BTZ black hole \cite{btz} in ``Schwarzschild'' coordinates is described
by the metric 
\begin{equation}  \label{a1}
ds^2=-(N^{\perp })^2dt^2+f^{-2}dr^2+r^2\left( d\phi +N^\phi dt\right) ^2
\end{equation}
with lapse and shift functions and radial metric 
\begin{eqnarray}  \label{a2}
N^{\perp } &=& f=\left( -M+{\frac{r^2}{\ell ^2}}+{\frac{J^2}{4r^2}}\right)
^{1/2}, \\
N^\phi &=& -{\frac J{2r^2}}\qquad (|J|\le M\ell ).  \nonumber
\end{eqnarray}
which satisfies the ordinary vacuum field equations of (2+1)-dimensional
general relativity (\ref{einst1}) with a cosmological constant $\Lambda
=-1/\ell ^2$.

If we are considering anholonomic frames, the matter fields ''deform'' such
solutions not only by presence of a energy--momentum tensor in the right
part of the Einstein equations but also via anisotropic polarizations of the
frame fields. In this Section we shall construct a subclass of d--metrics (%
\ref{dm3}) selecting by some particular distributions of matter energy
density $\rho (r)$ and pressure $P(r)$ solutions of type (\ref{a1}) but with
renormalized constants in (\ref{a2}), 
\begin{equation}  \label{renconst}
M\rightarrow \overline{M}=\alpha ^{(M)}M,J\rightarrow \overline{J}=\alpha
^{(J)}J,\Lambda \rightarrow \overline{\Lambda }=\alpha ^{(\Lambda )}\Lambda ,
\end{equation}
where the receptivities $\alpha ^{(M)},\alpha ^{(J)}$ and $\alpha ^{(\Lambda
)}$ are considered, for simplicity, to be constant (and defined
''experimentally'' or computed from a more general model of quantum 3D
gravity) and tending to a trivial unity value in the locally isotropic
limit. The d--metric generalizing (\ref{a1}) is stated in the from

\begin{equation}  \label{a3}
\delta s^2=-F\left( r\right) ^{-1}dt^2+F\left( r\right) dr^2+r^2\delta
\theta ^2
\end{equation}
where 
\[
F(r)=\left( -\overline{M}-\overline{\Lambda }r^2+{\frac{J^2}{4r^2}}\right),\
\delta \theta =d\theta +w_1dt\mbox{ and }w_1=-{\frac{\overline{J}}{2r^2}.} 
\]
The d--metric (\ref{a3}) is a static variant of d--metric (\ref{dm3}) when
the solution (\ref{faze}) is constructed for a particular function 
\[
\overline{\varpi }(r)=2\left( \frac{\kappa \overline{\rho }\left( r\right) }{%
r^2}-\overline{\Lambda }\right) 
\]
is defined by corresponding matter distribution $\overline{\rho }\left(
r\right) $ when the function $F(r)$ is the solution of equations (\ref{auxw2}%
) with coefficient $\overline{\varpi }(r)$ before $F^3,$ i. e. 
\[
FF^{\prime \prime }-(F^{\prime })^2+\varpi (r)F^3=0. 
\]

The d--metric (\ref{a3}) is singular when $r\!=\overline{\!r}_{\pm }$, where 
\begin{equation}  \label{rada}
\overline{r}_{\pm }^2=-{\frac{\overline{M}}{2\overline{\Lambda }}}\left\{
1\pm \left[ 1+\overline{\Lambda }\left( {\frac{\overline{J}}{\overline{M}}}%
\right) ^2\right] ^{1/2}\right\} ,
\end{equation}
i.e., 
\[
\overline{M}=-\overline{\Lambda }(\overline{r}_{+}^2+\overline{r}%
_{-}^2),\quad \overline{J}={\frac{2\overline{r}_{+}\overline{r}_{-}}{%
\overline{\ell }}}\ ,\overline{\Lambda }=-1/\overline{\ell }^2. 
\]

In locally isotropic gravity the surface gravity was computed \cite{Wald2} 
\[
\sigma ^2=-{\frac 12}D^\alpha \chi ^\beta D_\alpha \chi _\beta ={\frac{%
r_{+}^2-r_{-}^2}{\ell ^2r_{+}}}, 
\]
where the vector $\chi =\partial _v-N^\theta (r_{+})\partial _\theta $ is
orthogonal to the Killing horizon defined by the surface equation $%
r\!=\!r_{+}.$ For locally anisotropic renormalized (overlined) values we
have 
\[
\overline{\chi }=\delta _\nu =\partial _\nu -w_1(\overline{r}_{+})\partial
_\theta 
\]
and 
\[
\overline{\sigma }^2=-{\frac 12}D^\alpha \overline{\chi }^\beta D_\alpha 
\overline{\chi }_\beta =\overline{\Lambda }{\frac{\overline{r}_{-}^2-%
\overline{r}_{+}^2}{\overline{r}_{+}}}. 
\]

The renormalized values allow us to define a corresponding thermodynamics of
locally anisotropic black holes.

\subsection{Ellipsoidal black holes with running in time constants}

The anisotropic black hole solution of 3D vacuum Einstein equations (\ref%
{elipbh}) with elliptic horizon can be generalized for a case with varing in
time cosmological constant $\Lambda _0(t).$ For this class of solutions we
choose the local coordinates $(x^1=r,x^2=\theta ,y=t)$ and a d--metric of
type (\ref{dansatzc}), 
\begin{equation}  \label{dansatzc1}
\delta s^2 = \Omega _{(el)}^2(r,\theta ,t)[a(r)(dr)^2+b(r,\theta )(d\theta
)^2 +h(r,\theta ,t)(\delta t)^2],
\end{equation}
where 
\[
h(r,\theta ,t)=-\Omega _{(el)}^2(r,\theta ,t)=-\frac{4\Lambda ^3\left(
\theta \right) }{|\Lambda _0(t)|r^2}\left[ r_{+}^2(\theta ,t)-r^2\right] ^3, 
\]
for 
\begin{equation}
r_{+}^2(\theta ,t) = \frac{p(t)}{(1+\varepsilon \cos \theta )^2}, 
\mbox{ and
} p(t) = r_{+(0)}^2(\theta ,t)=-M_0/\Lambda _0(t)  \nonumber
\end{equation}
and it is considered that $\Lambda _0(t)\simeq \Lambda _0$ for static
configurations.

The d--metric (\ref{dansatzc1}) is a solution of 3D vacuum Einstein
equations if the 'elongated' differential 
\[
\delta t=dt+w_r(r,\theta ,t)dr+w_\theta (r,\theta ,t)d\theta 
\]
has the N--connection coefficients are computed following the condition (\ref%
{conformnc}), 
\[
w_r=\partial _r\ln |\ln \Omega _{(el)}|\mbox{ and }w_\theta =\partial
_\theta \ln |\ln \Omega _{(el)}|. 
\]

The functions $a(r)$ and $b(r,\theta )$ from (\ref{dansatzc1}) are arbitrary
ones of type (\ref{elips1}) satisfying the equations (\ref{vaceqel1}) which
in the static limit could be fixed to transform into static locally
anisotropic elliptic configurations. The time dependence of $\Lambda _0(t)$
has to be computed, for instance, from a higher dimension theory or from
experimental data.

\section{On the Thermodynamics of Anisotropic Black Ho\-les}

A general approach to the anisotropic black holes should be based on a kind
of nonequilibrium thermodynamics of such objects imbedded into locally
anisotropic gravitational (locally anisotropic ether) continuous, which is a
matter of further investigations (see the first works on the theory of
locally anisotropic kinetic processes and thermodynamics in curved spaces %
\cite{v4}).

In this Section, we explore the simplest type of locally anisotropic black
holes with anisotropically renormalized constants being in thermodynamic
equilibrium with the locally anisotropic spacetime ''bath'' for suitable
choices of N--connection coefficients. We do not yet understand the detailed
thermodynamic behavior of locally anisotropic black holes but believe one
could define their thermodynamics in the neighborhoods of some equilibrium
states when the horizons are locally anisotropically deformed and constant
with respect to an anholonomic frame.

In particular, for a class of BTZ like locally anisotropic spacetimes with
horizons radii (\ref{rada}) we can still use the first law of thermodynamics
to determine an entropy with respect to some fixed anholonomic bases (\ref%
{ddif}) and (\ref{dder}) (here we note that there are developed some
approaches even to the thermodynamics of usual BTZ black holes and that
uncertainty is to be transferred in our considerations, see discussions and
references in \cite{cm}).

In the approximation that the locally anisotropic spacetime receptivities $%
\alpha ^{(m)},\alpha ^{(J)}$ and $\alpha ^{(\Lambda )}$ do not depend on
coordinates we have similar formulas as in locally isotropic gravity for the
locally anisotropic black hole temperature at the boundary of a cavity of
radius $r_H,$%
\begin{equation}  \label{temp}
\overline{T}=-\frac{\overline{\sigma }}{2\pi \left( \overline{M}+\overline{%
\Lambda }r_H^2\right) ^{1/2}},
\end{equation}
and entropy 
\begin{equation}  \label{entropy}
\overline{S}=4\pi \overline{r}_{+}
\end{equation}
in Plank units.

For a elliptically deformed locally anisotropic black hole with the outer
horizon $r_{+}\left( \theta \right) $ given by the formula (\ref{elipsrad})
the Bekenstein--Hawking entropy, 
\[
S^{(a)}=\frac{L_{+}}{4G_{(gr)}^{(a)}}, 
\]
were 
\[
L_{+}=4\int\limits_0^{\pi /2}r_{+}\left( \theta \right) d\theta 
\]
is the length of ellipse's perimeter and $G_{(gr)}^{(a)}$ is the three
dimensional gravitational coupling constant in locally anisotropic media,
has the value 
\[
S^{(a)}=\frac{2p}{G_{(gr)}^{(a)}\sqrt{1-\varepsilon ^2}}arctg\sqrt{\frac{%
1-\varepsilon }{1+\varepsilon }}. 
\]
If the eccentricity vanishes, $\varepsilon =0,$ we obtain the locally
isotropic formula with $p$ being the radius of the horizon circumference,
but the constant $G_{(gr)}^{(a)}$ could be locally anisotropically
renormalized.

In dependence of dispersive or amplification character of locally
anisotropic ether with $\alpha ^{(m)},\alpha ^{(J)}$ and $\alpha ^{(\Lambda
)}$ being less or greater than unity we can obtain temperatures of locally
anisotropic black holes less or greater than that for the locally isotropic
limit. For example, we get anisotropic temperatures $T^{(a)}(\theta )$ if
locally anisotropic black holes with horizons of type (\ref{elipsrad}) are
considered.

If we adapt the Euclidean path integral formalism of Gibbon and Hawking \cite%
{gh} to locally anisotropic spacetimes, by performing calculations with
respect to an anholonomic frame, we develop a general approach to the
locally anisotropic black hole irreversible thermodynamics. For locally
anisotropic backgrounds with constant receptivities we obtain similar to %
\cite{btz94,cl,cm} but anisotropically renormalized formulas.

Let us consider the Euclidean variant of the d--metric (\ref{a3}) 
\begin{equation}  \label{a3e}
\delta s_E^2=\left( F_E\right) d\tau ^2+\left( F_E\right)
^{-1}dr^2+r^2\delta \theta ^2
\end{equation}
where $t=i\tau $ and the Euclidean lapse function is taken with locally
anisotropically renormalized constants, as in (\ref{renconst}) (for
simplicity, there is analyzed a non--rotating locally anisotropic black
hole), $F=\left( -\overline{M}-\overline{\Lambda }r^2\right),$ which leads
to the root $\overline{r}_{+}=\left[ -\overline{M}/\overline{\Lambda }\right]
^{1/2}.$ By applying the coordinate transforms 
\begin{eqnarray}  \label{coordt}
x & = & \left( 1-\left( \frac{{\overline r}_{+}}r\right) ^2\right) ^{1/2} \
\cos \left( -{\overline \Lambda} {\overline r}_{+}\tau \right) \exp \left( 
\sqrt{|{\overline \Lambda} |}{\overline r}_{+} \theta \right) ,  \nonumber \\
y & = & \left( 1-\left( \frac{{\overline r}_{+}}r\right) ^2\right) ^{1/2} \
\sin \left(-{\overline \Lambda} {\overline r}_{+} \tau \right) \exp \left( 
\sqrt{|{\overline \Lambda }|} {\overline r}_{+} \theta \right) ,  \nonumber
\\
z & = & \left( \left( \frac{{\overline r}_{+}}r\right) ^2- 1\right)
^{1/2}\exp \left(\sqrt{|{\overline \Lambda}|} {\overline r}_{+}\theta
\right) ,  \nonumber
\end{eqnarray}
the d--metric (\ref{a3e}) is rewritten in a standard upper half--space $%
\left( z>0\right) $ representation of locally anisotropic hyperbolic
3--space, 
\[
\delta s_E^2 = -\frac 1{\overline \Lambda }(z^2 dz^2+dy^2+\delta z^2). 
\]

The coordinate transform (\ref{coordt}) is non--singular at the $z$--axis $r=%
\overline{r}_{+}$ if we require the periodicity 
\[
\left( \theta ,\tau \right) \sim \left( \theta ,\tau +\overline{\beta }%
_0\right) 
\]
where 
\begin{equation}  \label{invtemp}
\overline{\beta }_0=\frac 1{\overline{T}_0}=-\frac{2\pi }{\overline{\Lambda }%
\ \overline{r}_{+}}
\end{equation}
is the inverse locally anisotropically renormalized temperature, see (\ref%
{temp}).

To get the locally anisotropically renormalized entropy from the Euclidean
locally anisotropic path integral we must define a locally anisotropic
extension of the grand canonical partition function 
\begin{equation}  \label{partfunct}
\overline{Z}=\int \left[ dg\right] e^{\overline{I}_E[g]},
\end{equation}
where $\overline{I}_E$ is the Euclidean locally anisotropic action. We
consider as for usual locally isotropic spaces the classical approximation $%
\overline{Z}\sim \exp \{\overline{I}_E[\overline{g}]\},$ where as the
extremal d--metric $\overline{g}$ is taken (\ref{a3e}). In (\ref{partfunct})
there are included boundary terms at $\overline{r}_{+}$ and $\infty $ (see
the basic conclusions and detailed discussions for the locally isotropic
case \cite{btz94,cl,cm} which are also true with respect to anholonomic
bases).

For an inverse locally anisotropic temperature $\overline{\beta }_0$ the
action from (\ref{partfunct}) is 
\[
\overline{I}_E[\overline{g}]=4\pi \overline{r}_{+}-\overline{\beta }_0M 
\]
which corresponds to the locally anisotropic entropy (\ref{entropy}) being a
locally anisotropic renormalization of the standard Bekenstein entropy.

\section{Chern--Simons Theories and Locally Anisotropic Gra\-vi\-ty}

In order to compute the first quantum corrections to the locally anisotropic
path integral (\ref{partfunct}), inverse locally anisotropic temperature (%
\ref{invtemp}) and locally anisotropic entropy (\ref{entropy}) we take the
advantage of the Chern--Simons formalism generalized for (2+1)--dimensional
locally anisotropic spacetimes.

By using the locally anisotropically renormalized cosmological constant $%
\overline{\Lambda }$ and adapting the Achucarro and Townsend \cite{at}
construction to anholonomic frames we can define two SO(2,1) gauge locally
anisotropic fields 
\[
A^{\underline{a}}=\omega ^{\underline{a}}+ \frac 1{\sqrt{\left| {\overline
\Lambda }\right| }} e^{\underline{a}} \mbox{ and } \widetilde{A}^{\underline{%
a}}=\omega ^{\underline{a}}-\frac 1{\sqrt{\left| {\overline \Lambda} \right| 
}}e^{\underline{a}} 
\]
where the index\underline{ }$\underline{a}$ enumerates an anholonomic triad $%
e^{\underline{a}}=e_\mu ^{\underline{a}}\delta x^\mu $ and $\omega ^{%
\underline{a}}=\frac 12\epsilon ^{\underline{a}\underline{b}\underline{c}%
}\omega _{\mu \underline{b}\underline{c}}\delta x^\mu $ is a spin
d--connection (d--spinor calculus is developed in \cite{v1}). The
first--order action for locally anisotropic gravity is written 
\begin{equation}  \label{act1}
\overline{I}_{grav}=\overline{I}_{CS}[A]-\overline{I}_{CS}[\widetilde{A}]
\end{equation}
with the Chern--Simons action for a (2+1)--dimensional vector bundle $%
\widetilde{E}$ provided with N--connection structure, 
\begin{equation}  \label{actcs}
\overline{I}_{CS}[A]=\frac{\overline{k}}{4\pi }\int\nolimits_{\widetilde{E}%
}Tr\left( A\land \delta A+\frac 23A\land A\land A\right)
\end{equation}
where the coupling constant $\overline{k}=\sqrt{\left| \overline{\Lambda }%
\right| }/(4\sqrt{2}G_{(gr)}) $ and $G_{(gr)}$ is the gravitational
constant. The one d--form from (\ref{actcs}) $A=A_\mu ^{\underline{a}}T_{%
\underline{a}}\delta x^\mu $ is a gauge d--field for a Lie algebra with
generators $\left\{ T_{\underline{a}}\right\} .$ Following \cite{cr} we
choose 
\[
\left( T_{\underline{a}}\right) _{\underline{b}}^{\quad \underline{c}%
}=-\epsilon _{\underline{a}\underline{b}\underline{d}}\eta ^{\underline{d}%
\underline{c}},~\eta _{\underline{a}\underline{b}}=diag\left( -1,1,1\right)
,~\epsilon _{\underline{0}\underline{1}\underline{2}}=1 
\]
and considering $Tr$ as the ordinary matrix trace we write 
\begin{eqnarray}
[T_{\underline{a}},T_{\underline{b}}]= f_{\underline{a}\underline{b}}^{\quad 
\underline{c}}T_{\underline{c}}= \epsilon _{\underline{a}\underline{b} 
\underline{d}}\eta ^{\underline{d}\underline{c}}T_{\underline{c}}, ~TrT_{%
\underline{a}}T_{\underline{b}}= 2\eta _{\underline{a}\underline{b}}, 
\nonumber \\
~g_{\mu \nu }= 2\eta _{\underline{a}\underline{b}}e_\mu ^{\underline{a}%
}e_\nu ^{\underline{b}},~\eta ^{\underline{a}\underline{d}} \eta ^{%
\underline{b}\underline{e}} f_{\underline{a}\underline{b}}^{\quad \underline{%
c}} f_{\underline{d}\underline{e}}^{\quad \underline{s}}= -2\eta ^{%
\underline{c}\underline{s}}.  \nonumber
\end{eqnarray}

If the manifold $\widetilde{E}$ is closed the action (\ref{act1}) is
invariant under locally anisotropic gauge transforms 
\[
\widetilde{A}\rightarrow A=q^{-1}\widetilde{A}q+q^{-1}\delta q. 
\]
This invariance is broken if $\widetilde{E}$ has a boundary $\partial 
\widetilde{E}.$ In this case we must add to (\ref{actcs}) a boundary term,
written in $\left( v,\theta \right) $--coordinates as 
\begin{equation}  \label{act2}
\overline{I}_{CS}^{\prime }=-\frac{\overline{k}}{4\pi }\int\nolimits_{%
\partial \widetilde{E}}TrA_\theta A_v,
\end{equation}
which results in a term proportional to the standard chiral
Wess--Zumino--Witten (WZW) action \cite{m,e}: 
\[
\left( \overline{I}_{CS}+\overline{I}_{CS}^{\prime }\right) [A]=\left( 
\overline{I}_{CS}+\overline{I}_{CS}^{\prime }\right) [\overline{A}]-%
\overline{k}\ \overline{I}_{WZW}^{+}[q,\overline{A}] 
\]
where 
\begin{equation}  \label{act3}
\overline{I}_{WZW}^{+}[q,\overline{A}]=\frac 1{4\pi }\int\nolimits_{\partial 
\widetilde{E}}Tr\left( q^{-1}\delta _\theta q\right) \left( q^{-1}\delta
_vq\right)
\end{equation}
\[
+\frac 1{2\pi }\int\nolimits_{\partial \widetilde{E}}Tr\left( q^{-1}\delta
_vq\right) \left( q^{-1}\overline{A}_\theta q\right) +\frac 1{12\pi
}\int\nolimits_{\widetilde{E}}Tr\left( q^{-1}\delta q\right) ^3. 
\]

With respect to a locally anisotropic base the gauge locally anisotropic
field satisfies standard boundary conditions 
\[
A_\theta ^{+}=A_v^{+}=\widetilde{A}_\theta ^{+}= \widetilde{A}_v^{+}=0. 
\]

By applying the action (\ref{act1}) with boundary terms (\ref{act2}) and (%
\ref{act3}) we can formulate a statistical mechanics approach to the
(2+1)--dimensional locally anisotropic black holes with locally
anisotropically renormalized constants when the locally anisotropic entropy
of the black hole can be computed as the logarithm of microscopic states at
the anisotropically deformed horizon. In this case the Carlip's results \cite%
{cr,gm} could be generalized for locally anisotropic black holes. We present
here the formulas for one--loop corrected locally anisotropic temperature (%
\ref{temp}) and locally anisotropic entropy (\ref{entropy}) 
\[
\overline{\beta }_0=-\frac \pi {4\overline{\Lambda }\hbar G_{(gr)}\ 
\overline{r}_{+}}\left( 1+\frac{8\hbar G_{(gr)}}{\sqrt{|\overline{\Lambda }|}%
}\right) \mbox{ and } \overline{S}^{(a)}=\frac{\pi \overline{r}_{+}}{2\hbar
G_{(gr)}}\left( 1+\frac{8\hbar G_{(gr)}}{\sqrt{|\overline{\Lambda }|}}%
\right) . 
\]
We do not yet have a general accepted approach even to the thermodynamics
and its statistical mechanics foundation of locally isotropic black holes
and this problem is not solved for locally anisotropic black holes for which
one should be associated a model of nonequilibrium thermodynamics.
Nevertheless, the formulas presented in this section allows us a calculation
of basic locally anisotropic thermodynamical values for equilibrium locally
anisotropic configurations by using locally anisotropically renormalized
constants.

\section{Conclusions and Discussion}

In this paper we have aimed to justify the use of moving frame method in
construction of metrics with generic local anisotropy, in general relativity
and its modifications for higher and lower dimension models \cite{v3,v4}.

We argued that the 3D gravity reformulated with respect to anholonomic
frames (with two holonomic and one anholonomic coordinates) admits new
classes of solutions of Einstein equations, in general, with nonvanishing
cosmological constants. Such black hole like and another type ones, with
deformed horizons, variation of constants and locally anisotropic
gravitational polarizations in the vacuum case induced by anholonomic moving
triads with associated nonlinear connection structure, or (in the presence
of 3D matter) by self--consistent distributions of matter energy density and
pressure and dreibein (3D moving frame) fields.

The solutions considered in the present paper have the following
properties:\ 1) they are exact solutions of 3D Einstein equations;\ 2) the
integration constants are to be found from boundary conditions and
compatibility with locally isotropic limits;\ 3) having been rewritten in
'pure' holonomic variables the 3D metrics are off--diagonal;\ 4) it is
induced a nontrivial torsion structure which vanishes in holonomic
coordinates;\ for vacuum solutions the 3D gravity is transformed into a
teleparallel theory;\ 5) such solutions are characterized by nontrivial
nonlinear connection curvature.

The arguments in this paper extend the results in the literature on the
black hole thermodynamics by elucidating the fundamental questions of
formulation of this theory for anholonomic gravitational systems with local
frame anisotropy. We computed the entropy and temperature of black holes
with elliptic horizons and/or with anisotropic variation and
renormalizations of constants.

We also showed that how the 3D gravity models with anholonomic constraints
can be transformed into effective Chern--Simons theories and following this
priority we computed the locally anisotropic quantum corrections for the
entropy and temperature of black holes.

Our results indicate that there exists a kind of universality of inducing
locally anisot\-rop\-ic interactions in physical theories formulated in
mixed holonomic--anholonomic variables:\ the spacetime geometry and
gravitational field are effectively polarized by imposed constraints which
could result in effective renormalization and running of interaction
constants.

Finally, we conclude that problem of definition of adequate systems of
reference for a prescribed type of symmetries of interactions could be of
nondynamical nature if we fix at the very beginning the class of admissible
frames and symmetries of solutions, but could be transformed into a
dynamical task if we deform symmetries (for instance, a circular horizon
into a elliptic one) and try to find self--consistently a corresponding
anholomic frame for which the metric is diagonal but with generic
anisotropic structure).

\subsection*{Acknowledgements}

The authors are grateful to D. Singleton, Radu Miron and M. Anastasiei for
help and collaboration. The S. V. work is supported both by a 2000--2001
California State University Legislative Award and a NATO/Portugal fellowship
grant at the Instituto Superior Tecnico, Lisboa.


\begin{thebibliography}{99}
\bibitem{at} A. Achucarro and P. K. Townsend, Phys. Lett. B {\bf 180,} 89
(1986).

\bibitem{barth} W. Barthel, J. Reine Angew. Math. {\bf 212,} 120 (1963).

\bibitem{btz} M. Ba\~nados, C. Teitelboim and J. Zanelli, Phys. Rev. Lett.\ 
{\bf 69}, 1849 (1992);\newline
M. Ba\~nados, M. Henneaux, C. Teitelboim and J. Zanelli, Phys. Rev. D\ {\bf %
48,} 1506 (1993).

\bibitem{btz94} M. Ba\~nados, C. Teitelboim and J. Zanelli, Phys. Rev.
Lett.\ {\bf 72}, 957 (1994).

\bibitem{cm} For reviews, see:\newline
S. Carlip, Class. Quantum Grav. {\bf 12,} 2853 (1995);\ R. B. Mann, ''Lower
Dimensional Black Holes: Inside and Out,'' Waterloo preprint
WATPHYS--TH--95--02, gr--qc/9501038 (1995).

\bibitem{cr} S. Carlip, Phys. Rev. D {\bf 51} 632 (1995).

\bibitem{cl} S. Carlip and C. Teitelboim, Phys. Rev. D {\bf 51} 622 (1995).

\bibitem{cartan1} E. Cartan, {\it La Methode du Repere Mobile, la Theories
des Groupes Continus et les Espaces Generalises} (Paris: Hermann, 1935).

\bibitem{chan} J. S. F. Chan, K. C. K. Chan and R. B. Mann, Phys. Rev. D 
{\bf 54,} 1535 (1996)

\bibitem{djh} S. Deser, R. Jackiv and G. 'tHooft, Ann. Phys.\ {\bf 52,} 220
(1984).

\bibitem{e} S. Elitzur et al., Nucl. Phys. {\bf B326,} 108 (1989).

\bibitem{haw} The energy conditions are discussed in: \newline
S. W. Hawking and C. F. R. Ellis, {\it The Large Scale Structure of Spacetime%
} (Cambridge University Press, 1973); \newline
K. V. Kuchar and C. G. Torre, Phys. Rev. D {\bf 43,} 419 (1991).

\bibitem{hus} V. Husain, Phys. Rev. D {\bf 53,} R1759 (1996);\  Phys. Rev. D 
{\bf 523,} (1995) 6860--6862;\ J. D. Brown and V. Husain, Int. J. Mod. Phys.
D 6 (1997) 563-573.

\bibitem{gh} G. W. Gibbons and S. W. Hawking, Phys. Rev. D {\bf 15}, 2752
(1977).

\bibitem{gm} A. Ghosh and P Mitra, Mod. Phys. Lett. A {\bf 11} (1996)
1231--1234.

\bibitem{Wald2} V. Iyer and R. M. Wald, Phys. Rev. D {\bf 52} (1995)
4430--4439.

\bibitem{kamke} E. Kamke, {\it Differential Gleichungen, Losungsmethoden und
Lonsungen: I. Gewohnliche Differentialgleichungen} (Lipzig, 1959).

\bibitem{kaw} T. Kawai, Phys. Rev. D {\bf 48,} 5668 (1993);\ Progr. Theor.
Phys. {\bf 94,} 1169 (1995),\ E--print:\ gr--qc/9410032.

\bibitem{ma} R. Miron and M. Anastasiei, {\it The Geometry of Lagrange
Spaces: Theory and Applications} (Kluwer Academic Publishers, Dordrecht,
Boston, London, 1994); {\it Vector Bundles. Lagrange Spaces. Applications in
Relativity} (Editura Academiei, Romania, 1987) [in Romanian] (English
translation: (Geometry Balkan Press, 1996)).

\bibitem{mtw} C. W. Misner, K. S. Thorne, and J. A. Wheeler, {\it \
Gravitation} (W. H. Freeman and Company, San Francisco, 1973).

\bibitem{m} G. Moore and N. Seiberg, Phys. Lett. B {\bf 220}, 422 (1989).

\bibitem{ross} S. Ross and R. B. Mann, Phys. Rev. D {\bf 47,} 3319 (1993).

\bibitem{v1} S. Vacaru, J. Math. Phys. {\bf 37,} 508 (1996); JHEP, {\bf 9809}
(1998) 011.

\bibitem{v2} S. Vacaru, Ann. Phys. (NY) {\bf 256,} 39 (1997); Nucl. Phys. 
{\bf B434,} 590 (1997); Phys. Lett. B {\bf 498}, 74 (2001).

\bibitem{v3} S. Vacaru, gr--qc/ 0001020; \ JHEP {\bf 0104,} 009\ (2001) {\sl %
\ }Ann. Phys. (NY) {\bf 290,} 83 (2001); S. Vacaru, et\ al., Phys. Lett. 
{\bf B 519,} 249 (2001); S. Vacaru and F. C. Popa, Class. Quant. Gravity. 
{\bf 18,} 4921 (2001); S. Vacaru and O. Tintareanu--Mircea, Nucl. Phys. B%
{\bf 626} (2002) 239; S. Vacaru and D. Singleton, J. Math. Phys. {\bf 43,}
2486 (2002); S. Vacaru and D. Singleton., Class. Quant. Gravity. {\bf 19,}
2793 (2002).

\bibitem{v4} S. Vacaru, Ann. Phys. (NY) {\bf 290,} 83 (2001);\ Ann. Phys.
(Leipzig) {\bf 11,} 5 (2000).

\bibitem{dw} B. S. DeWitt, in {\it Gravitation: An Introduction to Current
Research,} edited by L. Witten (Wiley, New York, 1962);\ Phys. Rev. {\bf 160,%
} 1113 (1967).

\bibitem{w} E. Witten, Nucl. Phys.\ {\bf B311,} 46 (1988).
\end{thebibliography}
\end{document}